\documentclass[5p,times]{elsarticle}
\usepackage{amsmath}
\usepackage{mathptmx}
\usepackage[cmintegrals]{newtxmath}
\usepackage[utf8]{inputenc}
\usepackage{graphicx}
\usepackage{color}
\usepackage{multirow}
\DeclareUnicodeCharacter{2212}{-}
\journal{Computer Networks}

\begin{document}

\begin{frontmatter}



\title{Intra-Body Communications for Nervous System Applications: Current Technologies and\\ Future Directions}


\author[inst1]{Anna Vizziello}

\affiliation[inst1]{organization={University of Pavia , \& Consorzio Nazionale Interuniversitario per le Telecomunicazioni},%Department and Organization
            addressline={Via Ferrata n. 5}, 
            city={Pavia},
            postcode={27100}, 
            %state={State One},
            country={Italy}}

\author[inst2]{Maurizio Magarini}
\author[inst1]{Pietro Savazzi}
\author[inst3]{Laura Galluccio}

\affiliation[inst2]{organization={Polytechnic University of Milan}, \& Consorzio Nazionale Interuniversitario per le Telecomunicazioni
            addressline={Piazza L. da Vinci 32}, 
            city={Milano},
            postcode={20133}, 
            country={Italy}}
\affiliation[inst3]{organization={University of Catania, \& Consorzio Nazionale Interuniversitario per le Telecomunicazioni},%Department of Electric Electronic and Computer Engineering
            addressline={V.le A. Doria 6}, 
            city={Catania},
            postcode={95125}, 
            country={Italy}}            
\graphicspath{{Figures/}}

\begin{abstract}
The Internet of Medical Things (IoMT) paradigm will enable next generation healthcare by enhancing human abilities, supporting continuous body monitoring and  restoring lost physiological functions due to serious impairments.
\textcolor{black}{This paper presents intra-body communication solutions that interconnect implantable devices for application to the nervous system, challenging the specific features of the complex intra-body scenario.  The presented approaches include both speculative and implementative methods, ranging from neural signal transmission to testbeds,  to be applied to specific  neural diseases therapies. Also future directions in this research area are considered  to overcome the existing technical challenges mainly associated with miniaturization, power supply, and multi-scale communications.
}
\end{abstract}


%\item \textcolor{red}{Research highlight 1}
%\item \textcolor{red}{Research highlight 2}

\begin{keyword}
Internet of things \sep nervous system applications \sep electromagnetic communications \sep coupling
technologies \sep ultrasounds \sep experimental testbeds \sep intra-body networks 
%\PACS \textcolor{red}{0000} \sep \textcolor{red}{1111}
%\MSC \textcolor{red}{0000} \sep \textcolor{red}{1111}
\end{keyword}

\end{frontmatter}

\section{Introduction}
\label{Intro}
In recent decades, Implantable Medical Devices (IMDs) have become a reality with the progress in engineering technologies that include nano-materials, micro and nanoelectronics, as well as biotechnology. A dominant market for them is North America, where 10$\%$ of the population employ IMDs for several reasons: diagnosis, therapy, and assistive technical purposes~\cite{AMR}.
%~\cite{IMDoptoMag2020}.
Also, the Asia-Pacific market is experiencing a fast growth with consequent improvements of healthcare policies~\cite{AMR}.

Some solutions have been already implemented, such as deep brain stimulators, glucose sensors, and heart pacemakers. 
This will promote next generation healthcare by enabling personalized  medicine through real-time physiological monitoring and proactive drug delivery, as envisioned in the fifth and beyond fifth generation (5G/B5G) communication scenarios~\cite{IEEETMBMC2020}.
\textcolor{black}{However, more challenging applications can be designed, for which the current single transmission between an implant and an external monitoring center would not be sufficient. Rather, it would be necessary to ensure the communication among  implants, requiring the development of body-centric architectures for the exchange of information inside/outside the body~\cite{dwivedi2021potential,srivastava2022internet}. These solutions may leverage both intranet and Internet of Medical Things (IoMT) architectures.}

\begin{figure}%[!t]
    \centering
	\includegraphics[width=0.4\textwidth]{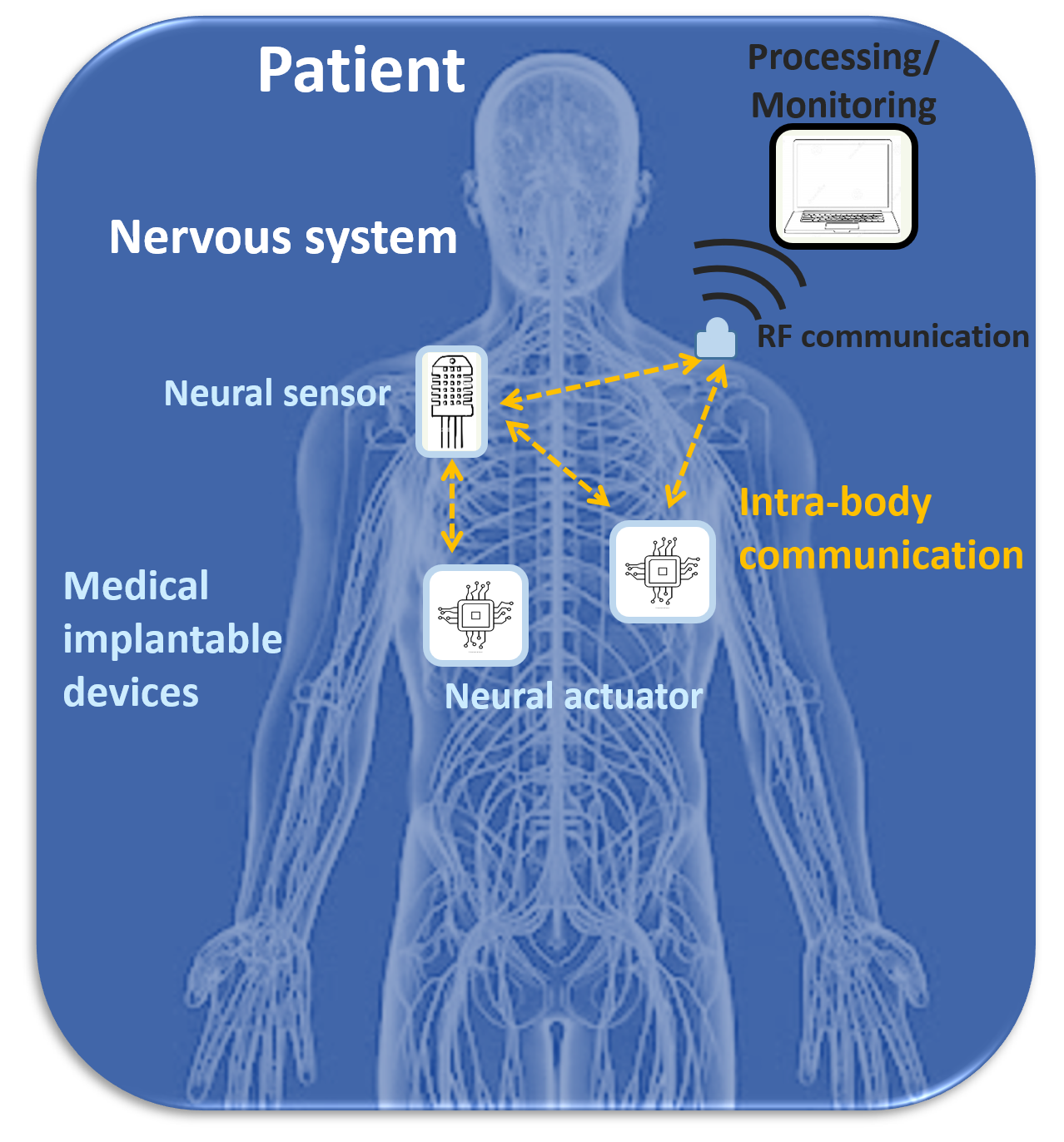}
	\caption{Nervous System Application Scenario using Intra-Body Communication for sensing and delivering biological data.} \label{fig:NervousSystemImplants}
\end{figure}

One of the most visionary applications is related to the use of devices able to interact with cells of the nervous system either at a microscopic or at a macroscopic level, to recover motor function or bypass spinal cord damages due to serious injuries (Fig.~\ref{fig:NervousSystemImplants}). 
The support for these disruptive applications, however, calls for the possibility to set up a communication network among devices. 
Indeed, the functional recovery of damaged neurons can be enhanced by means of innovative implantable sensor and actuator devices that exchange information.

\begin{figure}[!t]
\centering
	\includegraphics[width=.45 \textwidth]{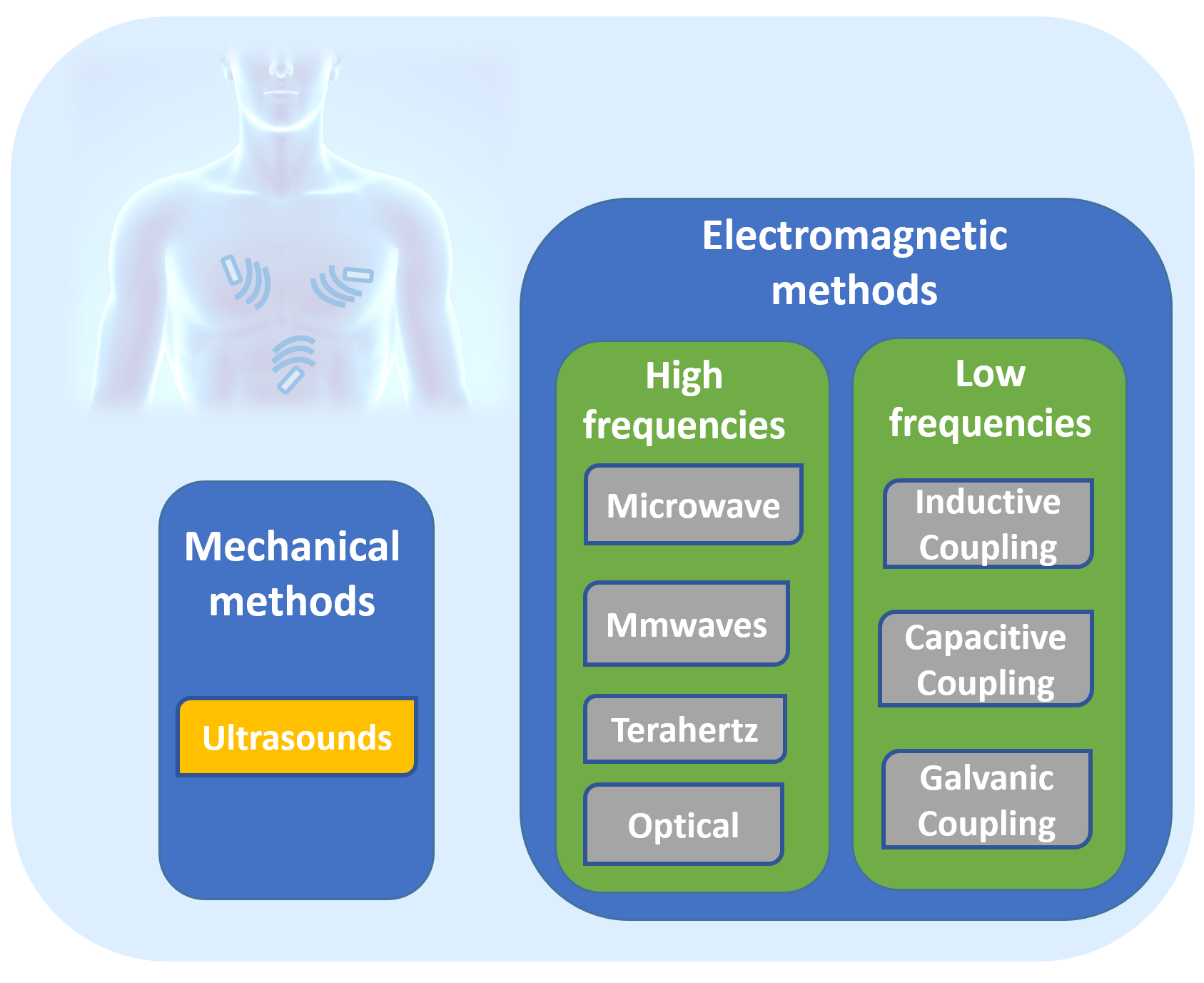}
	\caption{Taxonomy of intra-body communication technologies.} \label{fig:Tech}
\end{figure}

\textcolor{black}{The intra-body network paradigm enables the interconnection} of devices inside the body by enabling the transmission of the sensed measurements among them or to an external monitoring center. %for processing  by means of an Internet of Things (IoT) architecture, or also to receive updates and commands for the embedded actuators. 
\textcolor{black}{To this aim, it is necessary to design energy efficient communication techniques.
The most common intra-body communication method uses classical Radio Frequency (RF) waves, working at frequencies below $1\,$GHz or at the Industrial, Scientific and Medical (ISM) band, in the form of narrowband (NB) or ultra-wideband (UWB) signals~\cite{Cavallari2014}.} %for communication between implants or between an implant and a relay placed on body~\cite{Cavallari2014}.
However, several studies have demonstrated that RF signals, although profitably used for the communication of on body wearable devices, undergo high losses within the tissue~\cite{GC1}, with consequent short distance coverage and \textcolor{black}{potential tissue heating that could damage it. Hence, other technologies have been explored as profitable alternatives for intra-body communication, %which have unique properties that may be more suited, 
including ultrasounds (US) and electromagnetic (EM)-based methods~\cite{MC1,LauraBV,GC7,mmW1,OPTO,THzAlouini2022} (Fig.~\ref{fig:Tech}).}
%working at low or high frequencies~\cite{MC1,LauraBV,GC7,mmW1,OPTO,THzAlouini2022} (Fig.~\ref{fig:Tech}). %\cite{MC1}-\cite{OPTO}.

\textcolor{black}{This article discusses for the first time holistically the applicability of these non-traditional communication methodologies to the complex and challenging nervous system. Interesting works in the past presented engineered solutions for nervous system applications; however, their biomedical engineering view is different from the communication perspective presented here, since they mainly focus on brain-machine interfaces (BMI), exoskeletons, or neuroprosthetics~\cite{Gheng2020, Raspopovic2020, Das2020}. }

\textcolor{black}{Differently, we focus on solutions that specifically exploit wireless communication technologies inside the human body for nervous system applications. To this purpose, speculative studies on neural communication in the context of molecular communications (MC) were conducted, which provide great insight into the realm of nervous system knowledge. These works leverage ICT based communication methods and approaches to analyze the biological neural communication system  ~\cite{MCMladen,MC4,MC5,MC6,MC7,MC8,MC9,MC10}. Some papers discussed MC as a promising nano-technology for human health applications, such as~\cite{MC1} that reports a preliminary experimental platform~\cite{MC2} for nervous system applications. 
However, the considered nano-devices for communication are still at the design stage and the experimental platforms at an early level. 
Therefore, in a complementary way to MC, in this paper we specifically focus on intra-body communication technologies that achieve a higher Technology Readiness Level (TRL) and, unlike RF solutions, show good intra-body communication properties when used inside the body tissue. 
The considered technologies are US, EM methods working at low frequencies, which are based on inductive (IC), capacitive (CC), and galvanic coupling (GC), and EM methods working at higher frequencies, i.e., microwave, millimeter waves (mmWaves), Terahertz (THz), and optical frequencies (Fig.~\ref{fig:Tech}). 
%{\bf The contribution of the aforementioned works provide great insight into the realm of nervous system applications.???togliere} 
Furthermore, noting that there is no single work in which all of the above intra-body technologies are discussed holistically and compared in terms of their ability to transmit neural signals, their application to neural diseases, and the existing testbeds in this application area, we fill this important gap. 
%Therefore, in this article we review a number of solutions where we investigate the use of wireless communication among implantable devices. 
Finally, we highlight future directions for miniaturization of neural implantable devices, discussing in particular the communication challenges.}

\textcolor{black}{The article is organized as follows. After a description about intra-body communication technologies in Sec. \ref{IntraCommMethods}, their applications to the nervous system are detailed in Sec. \ref{MacroApp}. The existing testbed specifically designed for nervous system applications are detailed in Sec. \ref{SecTestbeds}, while future directions in this research area and conclusions are finally presented in Sec. \ref{Directions} and Sec.\ref{Conclusions}.} 
%{\bf vanno messe le sottosez?}

\section{Background on Intra-Body Communication Methods}
\label{IntraCommMethods}

\begin{figure}[!t]
\centering
	\includegraphics[width=.45\textwidth]{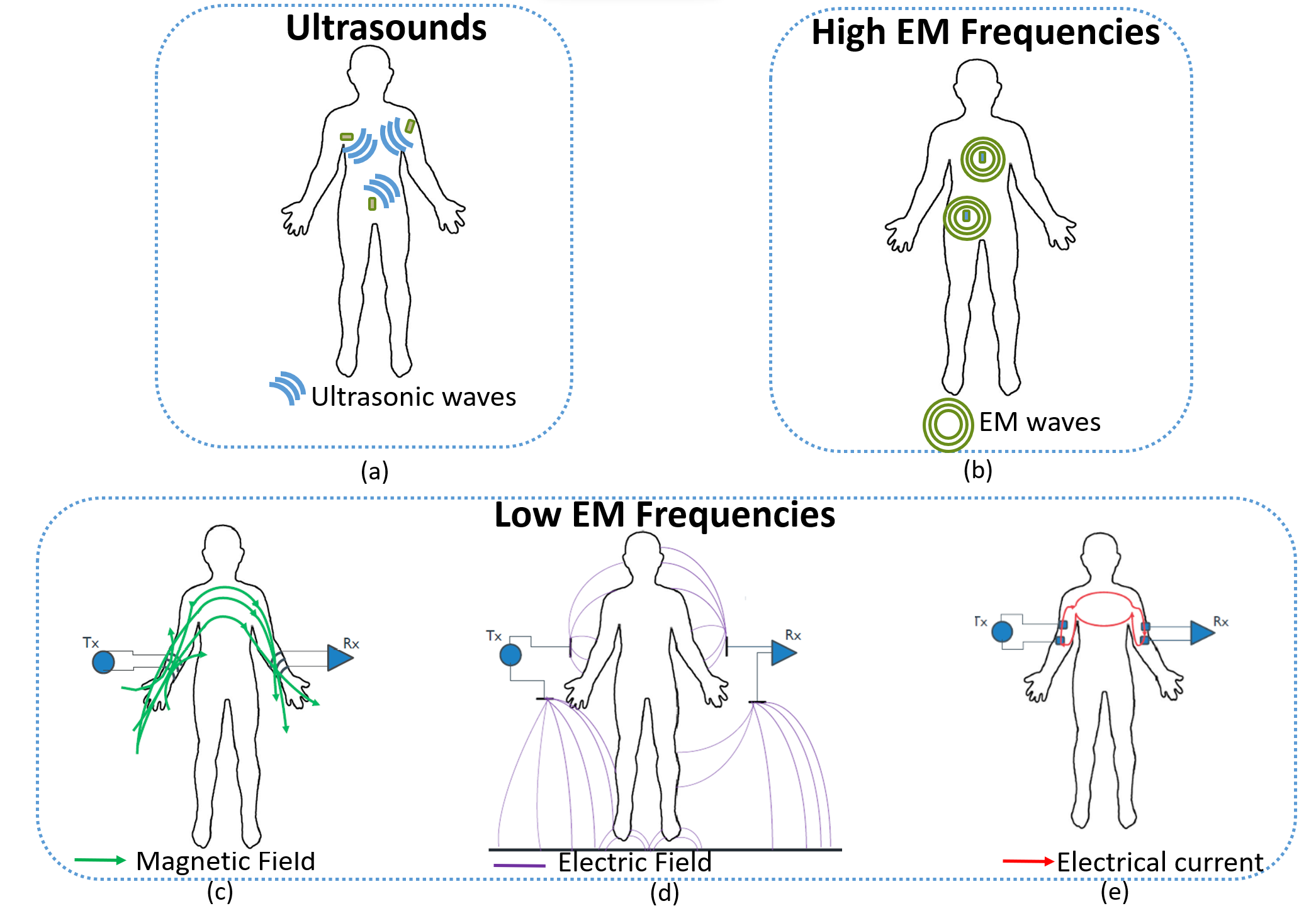}
	\caption{\textcolor{black}{Physical principle of (a) US (b) high EM frequencies (c) IC (d) CC (e) GC ~\cite{GC1}.} }\label{fig:PHYtechs}
\end{figure}

\textcolor{black}{In this section, we present some fundamentals on intra-body communication technologies. %as an alternative to the traditional RF method. %that can be used to support nervous system applications. 
%Indeed, given the high signal losses of traditional RF solutions when used inside the body, we overview use of alternative low-power technologies for transmission inside or on the body. %(Fig.~\ref{fig:Tech}). 
These include US, which use mechanical vibrations, and EM methods,} which may work both at low or high frequencies (Fig.~\ref{fig:Tech}). Their physical principles are summarized in Fig.~\ref{fig:PHYtechs}~\cite{GC1}.

\subsection{Ultrasounds}
%\textcolor{red}{@ANNA: HO TAGLIATO QUALCOSA MA SE GUARDI L'ESTENSIONE RISPETTO AD HF E LF NON E' MOLTO DIVERSO COME ESTENSIONE quindi lascerei cosi'}

US are acoustic waves (Fig.~\ref{fig:PHYtechs}(a)) generated by an electrical driving signal that is transduced into a vibrational mechanical wave above $20\,$kHz~\cite{US2} (the human audible limit). Considering the unique features of human body tissues composed of more than 65\% of water, it was observed that frequencies in the range from $700\,$kHz to $5\,$MHz allow to achieve an optimal trade-off between attenuation, directivity, and coverage range extension~\cite{ref7, ref17}. \textcolor{black}{In ultrasonic propagation,  multipath fading is met \cite{ref8}  due to reflections and refraction. 
The absorption is very low without any damage to tissues, as evidenced by the safe use of US in medicine for diagnosis and therapy over the past six decades}. 

\textcolor{black}{Intra-body ultrasonic technology was proved to be a possible approach for body area network communications while also allowing to both send recorded neural data (i.e. using US to perform  data transmission) and do harvesting~\cite{144Das2020} (i.e. transfer power by means of US). }
Also, \textcolor{black}{well-established} usage of US for pipe-integrity analysis is the basis of the recent experimental evidence that showed how ultrasonic waves tunnel over intra-body fluid tubes, specifically blood vessels, can be employed to remarkably increase the communications range inside the human body~\cite{LauraBV}. Experimental results assess the effectiveness of using diffusive ultrasonic body communications by exploiting multiple hops among network nodes to extend the \textcolor{black}{single-hop} communication range usually limited around $15\,$cm~\cite{ref8}.

\textcolor{black}{This limited communication range, typical of diffusive ultrasonic communications, results from the challenging propagation of ultrasonic waves inside the human body. Indeed, possible reflections, e.g., at the boundary between muscle and bones, or scattering against tissues and particles can occur, as investigated  in ~\cite{ref8}. 
The usage of wave guided ultrasonic propagation inside blood vessels~\cite{LauraBV} allows} to extend the \textcolor{black}{single-hop} communication range to approximately $35-40\,$cm~\cite{LauraBV}. 

\textcolor{black}{Regarding ultrasonic nano-transducers manufacturing, multiple technologies have been proposed to be employed in recent years. On the one hand, piezoelectric micromachined ultrasonic transducers (pMUTs)~\cite{ref26,PMUT} use MEMS for piezo actuation; on the other hand, capacitive micromachined ultrasonic transducers (cMUTs)~\cite{ref26} consist of two electrodes, a top and a bottom one, and exploit energy transduction due to change in capacitance.} A cavity is formed in a silicon substrate, and a thin layer suspended on the top of the cavity serves as a membrane on which a metallized layer acts an electrode. If a signal is applied across the biased electrodes, the vibrating membrane will produce ultrasonic waves in the medium of interest. \textcolor{black}{As an alternative, opto-ultrasonic transducers~\cite{US2} have been proposed, which rely on the opto-acoustic effect to generate acoustic waves by exploiting light absorption on a material. Finally, \textcolor{black}{single-layer} graphene (SLG)~\cite{ref28} devices may also represent an alternative.} When an electric signal of sound frequency is applied to SLG devices, the air near its surface will be heated by Joule heating. Then, the periodic vibration of the air will  lead to formation of sound waves.

When US waves propagate through an absorbing medium, the initial pressure \textcolor{black}{decays} as a function of the distance from the source according to an exponentially decreasing function of different parameters, such as carrier frequency and tissue attenuation. Since most intra-body sensing applications require highly directional transducers, it is needed to operate at high frequencies to keep the transducer size small, while also considering the beam directivity. Conversely, higher transmission frequency leads to higher attenuation. For distances higher than a few cm, it has been observed that the transmission frequency should not exceed $10\,$MHz. 

%Concerning directivity issues, depending on the desired application, transducers can be designed to radiate sound according to different patterns depending on the ratio of the diameter of the radiating surface and the wavelength at the operating frequency~\cite{ref7}.

%Regarding noise, in the range of the US frequencies only the thermal noise is relevant. In our previous studies~\cite{ref8} we characterized the channel response as a sum of three contributions, related to the direct, the reflected, and the lateral components which propagate on the surface of the body and that are combined due to multipath.  

\subsection{High EM Frequencies}

\textcolor{black}{Microwaves, mmWaves, higher THz bands, and optical bands are included among these techniques. They have been grouped together in Fig.~\ref{fig:PHYtechs}(b) because they leverage on the same working principle, although the covered distance, tissue penetration, and tissue channel behaviour depend on the specific EM frequencies used.}
%Among technologies working at high EM frequencies, 
%Microwaves have been employed to transmit data within the fat tissues as in the FET Project B-CRATOS. 
%{\bf with high bandwidth???che vuol dire}.

Microwave based solutions working around 2 GHz have been developed, with the specific feature to  utilize the fat tissue as channel for data transmission, \textcolor{black}{and was named fat-microwave} ~\cite{FAT1,FAT2,FAT3,FAT4,FAT5,FAT6}. %\textcolor{red}{(ANNA: mettere nelle ref i paper della tesi p.5/118)}. 
\textcolor{black}{At these high frequencies, large bandwidth (2 MHz) is allowed when transmitting across  fat tissues channels, thus exhibiting good communication properties compared to the case of skin and muscle tissues. %Accordingly, microwave propagation is confined to the subdermal body fat with minimal interference from external electronic implantable devices. 
Indeed, the fat tissue layer is located between the skin and the muscle layers forming a type of parallel plate waveguide having two conductors on both sides of a dielectric material.}
The high contrast in dielectric properties between fat and skin, as well as between fat and muscle, allows the signal to be confined within the fat layer that acts as a waveguide, hence a low-loss communication medium. This interesting method may potentially increase the transmission range achievable at microwaves~\cite{FAT3}. \textcolor{black}{Anyhow, it is worth to note that this configuration is not perfectly equivalent to a parallel waveguide since the skin and muscle are lossy media, which turns out in not perfect conductor boundary conditions~\cite{FAT3}. Both simulations and experiments were conducted in presence of perturbants, such as embedded muscle layers and blood vessels. The results showed that the transmission method is feasibile even in those conditions since the communication channel is not affected by perturbants smaller than $15\,$mm$^3$~\cite{FAT5}.} Preliminary experimental results were conducted with a  multi-layer tissues phantom achieving a data rate around $250\,$kb/s and maximum $25\,$dB attenuation with a $10\,$cm coverage distance~\cite{FAT3} - even if distances up to one meter are expected~\cite{FAT2}.

%\textcolor{black}{In \cite{Yao-Hong2022IR-UWB} implantable impulse-radio ultra-wideband (IR-UWB) wireless systems, operating in a 6–9-GHz UWB band, are considered for intracortical neural sensing interfaces. These technologies allow a high energy efficiency of 5.8 pJ/bit, and a high data rate of 1.66 Gbps, obtained through 8-PSK and 4-PPM.}

\textcolor{black}{In \cite{Yao-Hong2022IR-UWB}, an impulse-radio ultra-wideband (IR-UWB) wireless telemetry system was developed, capable of achieving a high data rate in the order of Gbps %of 1.66 Gbps 
in the 6–9-GHz UWB band.}

Even larger bandwidth may be obtained at the millimeter waves, as tested for on-body/in-body data transmission~\cite{mmW1,mmW2}. However, studies have revealed that at $60\,$GHz around 40$\,\%$ of the incident power is reflected at the skin surface and $90\,\%$ of the signal is absorbed within the skin with consequent tissue heating~\cite{mmW1}. Hence, only wearable configurations can be considered as viable applications reaching up to $50\,$cm with an operating frequency range $30-300\,$GHz and a data rate in the order of Gb/s~\cite{mmW2}.

%FRASE PLASMONIC ANTENNA PRESA da Sec. II.B di ~\cite{IEEETMBMC2020}
Concerning higher frequencies, THz ($0.1-10\,$THz)~\cite{NanoTHz1}, near-infrared ($300-400\,$THz)~\cite{OPTO1,OPTO2}, and optical frequencies  ($400-750\,$THz)~\cite{OPTO1,OPTO2,OPTO3,OPTO4} have been explored for communication among implants thanks to the possibility of employing the recently developed idea of plasmonic nano-antennas for wireless optical communication~\cite{OPTO2}. Indeed, those frequencies require nano-devices and nano-antennas, which can not be developed by exploiting the classical metallic antenna technology. To minimize the metallic antenna size, the resonance frequencies should be very high, for example 1$\mu$m long dipole antenna will resonate at $150\,$THz~\cite{IEEETMBMC2020}. \textcolor{black}{However, in
this area, metals do not behave as perfect electric conductors anymore~\cite{IEEETMBMC2020}. Anyhow, with the development of new materials showing plasmonic effects, such as graphene at THz band~\cite{IEEETMBMC2020} and noble metals and metamaterials at optical frequencies~\cite{OPTO1}}, nano-transceivers and nano-antenna design become possible. A plasmonic nano-antenna was proposed for the first time by using graphene where a $1\,$mm long nano-antenna can radiate at frequencies in the THz band~\cite{grapheneAntenna}.

Anyhow, at these high frequencies much shorter transmission distances can be covered (see Table \ref{TabTechs}) due to several phenomena, such as the absorption and scattering of different type of cells, which challenge electromagnetic waves' propagation inside the body~\cite{OPTO} and can not be captured with traditional channel models developed for lower frequencies. The classical multi-layer approximation of the human tissues with different permeabilities and permittivities is not applicable at these higher frequencies where the body becomes rather a collection of several types of elements, such as biological cells and molecules, with different geometry, placement and electromagnetic properties. Hence, the absorption and scattering of different types of cells need to be considered for data communication at these frequencies. Thus, appropriate channel models have been developed at these frequencies, such as~\cite{THzChannel1,THzChannel2,THzChannel3}. %\textcolor{red}{(references channel model in Sec. II.B pag. 4 di \cite{IEEETMBMC2020}, loro ref 54,55,56)}
Furthermore, concerning the THz band, the absorption of liquid water molecules challenges the propagation of the waves inside the human body. 
\textcolor{black}{Although THz band radiation is not ionizing, which means that it can not cause any damage to the molecular structure, the propagation of THz-band waves inside the human body is drastically impacted by the absorption of liquid water molecules 
\cite{Thz-Opto-LinkBudget}.
Hence, the use of the optical window has been advocated \cite{OPTO1} since absorption from liquid water molecules is minimal in the so-called optical window (from 400 THz and 750 THz) \cite{Thz-Opto-LinkBudget}. Moreover, plasmonic nanodevices at optical frequencies have already been used in several in vivo applications \cite{Thz-Opto-LinkBudget}.}

%MajorRev, da \cite{Thz-Opto-LinkBudget}: As a result, Guo et al. \cite{OPTO1}, advocated the use of the optical window for intrabody wireless communication among nanosensors with plasmonic nanoantennas. This is due to the fact that the absorption from liquid water molecules is minimal in the so-called optical window, roughly between 400 THz and 750 THz [12]. In addition, plasmonic nanodevices at optical frequencies have already been utilized in several in vivo applications [13].

\subsection{Low EM Frequencies} 
\textcolor{black}{Technologies at low frequencies are usually classified based on their coupling principle. They use different physical methods to generate an electrical signal that propagates through the human body.} The electrical signal is below $200\,$MHz with low-power (in the order of $\mu$W) compared to traditional RF signals working up to several GHz~\cite{GC0}. For this reason, coupling methods have gained great attention in the on-going research on intra-body communication ensuring safety and decreasing energy consumption.
They are classified in IC, CC and GC techniques~\cite{GC7}.

In the first coupling mechanism, magnetic energy is generated and received by means of coils (Fig.~\ref{fig:PHYtechs}c). Indeed, it is possible to configure two wires so that they are inductively coupled, and hence a current change in one of them induces a voltage across the other by way of EM induction. The coupling between the two wires can be enhanced by twisting them into coils and putting them close each other~\cite{GC6}. This technology is largely employed for wireless power transfer to implanted devices up to few centimeters, but also to transfer data typically operating below $10\,$MHz. However, the transmission efficiency is limited by the coupling efficiency associated with the resonance frequency matching between the transmitter and the receiver, which is often difficult to be achieved.

CC and GC show some similarities since both of them utilize electrodes as transmitter and receiver, although in different configurations. In CC, only one of the two transmitter electrodes is attached to the body because the other (ground electrode) floats. The same setup is used for the receiver. The physical principle is based on the near-field electrostatic coupling of the human body with its surrounding environment (Fig.~\ref{fig:PHYtechs}d). %The body acts as a conductor and the signal is confined by surface waves on the direct path formed through the human body while the external ground behaves as a return path~\cite{GC6}. 
The carrier frequency ranges from $100\,$kHz up to $100\,$MHz~\cite{GC0,GCPastFuture}, and can be used for wearable scenarios covering long distances up to $170\,$cm, although it may affected by environmental conditions.

In GC both the pairs of transmitter and receiver electrodes are attached to or implanted in the body. \textcolor{black}{While the primary current %carrying the data 
flows between the two transmitting electrodes, low secondary currents move away from the transmitter and can be detected at the receiver electrodes} (Fig.~\ref{fig:PHYtechs}e). This technology is suitable for implanted scenarios and consumes two orders of magnitude less energy than RF~\cite{GC8}. Its operating frequency is in the range $1\,$kHz-$100\,$MHz (to not impair other natural signals with a transmission distance of $20-30\,$cm~\cite{GC7}) with a coverage range of tens of cm. \textcolor{black}{The frequency channel response is relatively flat in the range of interest}~\cite{GCch1,GCch2}, which is suitable to design simple implanted transceivers as required in intra-body networks.

%\textcolor{black}{A transmission method based on digital-impulse GC for high-speed trans-dural (from cortex to the skull) data transmission has been proposed in \cite{Yao-Hong2022GC}. This technology allows the brain implant to be “free-floating” for minimizing brain tissue damage, showing that the trans-dural channel has a wide frequency response of up to 250 MHz. In \cite{Yao-Hong2022Sensing}, a bio-inspired, event-driven neuromorphic sensing system (NSS) has been presented, considering an on-chip feature extraction and “send-on-delta” pulse-based transmission for peripheral nerve neural recording applications.}
%\textcolor{black}{In \cite{Yao-Hong2022GC} a new as a new high-speed trans-dural data transmission, based on GC, is presented The proposed in order to replace the tethered wires connected in between implants on the cortex and above the skull, allowing the brain implant to be “free-floating” for minimizing brain tissue damage.}

%\textcolor{red}{aggiunto capacitivo impiantato che oltre alla teoria, sarebbe implementazione di IBCOM \cite{GC3}:}
In summary, GC is usually employed for communication between implanted devices, while CC  for establishing communication between on-body devices or devices very close to the body. However, it has been recently demonstrated that a stable capacitive return path can be obtained not only by exposing the capacitive ground electrode directly to the air, which is feasible in wearable configurations, but also in implantable devices, provided that the ground electrode is isolated from the human tissue~\cite{CCimplanted1,CCimplanted2,CCimplanted3,CCimplanted4}. Therefore, capacitive intra-body method emerges as a viable alternative for communication with implanted devices that can enhance the achievable transmission range inside the body. The results of implantable CC are very promising but this area is still in a nascent stage, for example a deep channel characterization of this configuration is still missing and further investigation is required~\cite{CCimplanted3}.

\subsection{\textcolor{black}{Trade-off comparison among intra-body communication technologies}}

\begin{figure}[!t]
    \centering
	\includegraphics[width=0.45\textwidth]{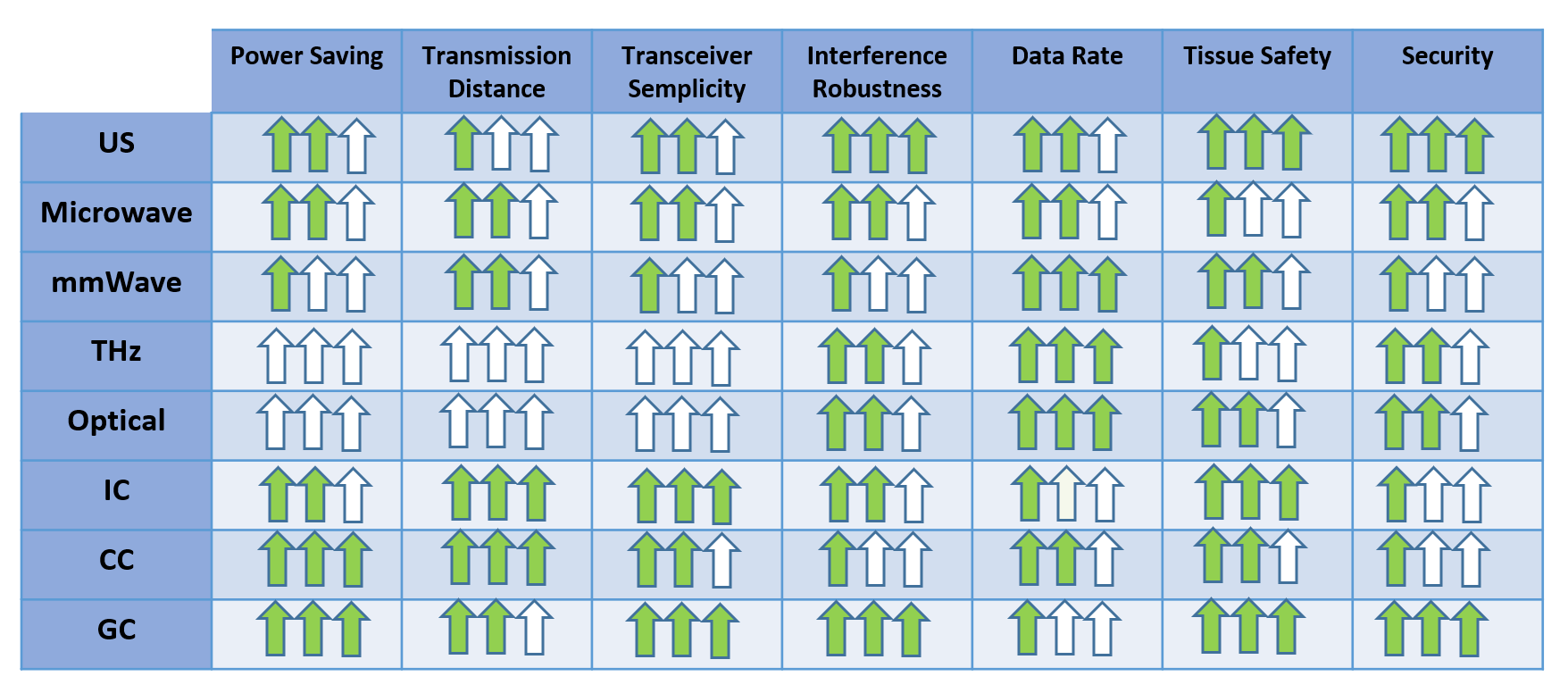}
	\caption{\textcolor{black}{Comparison of transceiver design constraints for different intra-body communication technologies.}} \label{fig:Fig_compare}
\end{figure}

\textcolor{black}{We compare the intra-body communication methods here, in terms of their channel characteristics (Table~\ref{TabChannelTechs}) and of their main physical layer features (Table~\ref{TabTechs}). Moreover, a qualitative comparison of transceiver design constraints is presented in Fig.~\ref{fig:Fig_compare}.}

\begin{table}[!t] \footnotesize
    \renewcommand{\arraystretch}{1.1}
    \caption{\textcolor{black}{Comparison of channel characteristics for different intra-body communication methods. US refers to ultrasounds, IC to inductive coupling, CC to capacitive coupling and GC to galvanic coupling.}}
    \centering
    \begin{tabular}{ccc} 
    \hline
          \textcolor{black}{\multirow{2}{*}{\textbf{Technology}}}&\textcolor{black}{\textbf{Channel Model}}&\textcolor{black}{\textbf{Channel}}\\
          &\textcolor{black}{\textbf{Method}}&\textcolor{black}{\textbf{Characteristics}}\\
        \hline
        \textcolor{black}{\multirow{2}{*}{US \cite{ref8}}} & \textcolor{black}{Analytical and}& \textcolor{black}{\multirow{2}{*}{Multipath Fading}}\\
        &\textcolor{black}{Empirical}&\\
        \\
        \textcolor{black}{Microwave \cite{FAT5}} &\textcolor{black}{Numerical} &\textcolor{black}{Frequency Selective}\\
        \\
        \textcolor{black}{mmWave \cite{mmW2}} &\textcolor{black}{Statistical} &\textcolor{black}{Cauchy-Lorenz Fading}\\
        \\
        \textcolor{black}{\multirow{2}{*}{THz \cite{Thz-Opto-LinkBudget}}} &\multirow{2}{*}{\textcolor{black}{Analytical}}&
        \textcolor{black}{Molecular Absorption,}\\
        &&\textcolor{black}{Spreading, Scattering}\\
        \\
         \textcolor{black}{Optical \cite{OPTO1}} &\textcolor{black}{Analytical}&
         \textcolor{black}{Multiple Scattering Model}\\%\textcolor{black}{Scattering from cells} \textcolor{red}{OPPURE}\\
        \\
        \textcolor{black}{IC \cite{ICTable}} &\textcolor{black}{Empirical}&\textcolor{black}{Frequency Selective}\\
        \\
        \textcolor{black}{CC \cite{CCwang_ch}} &\textcolor{black}{Empirical}&\textcolor{black}{Frequency Selective}\\
        \\
        \textcolor{black}{GC \cite{GCch1}} &\textcolor{black}{Empirical}&\textcolor{black}{AWGN}\\
        \hline
    \end{tabular}\label{TabChannelTechs}
\end{table}

\textcolor{black}{Table \ref{TabChannelTechs} illustrates the channel modeling methods employed to characterize channel behavior for all the aforementioned technologies. Several approaches are available in literature, i.e., empirical, analytical, statistical, and numerical methods \cite{FAT5,mmW2,OPTO1,THzChannel2,GCch1,Guan_USch,ICTable,CCwang_ch}. However, most works use a computational based method supplemented by empirical validation. 
%Table \ref{TabChannelTechs} shows that GC offers the most simplistic channel behavior, while other methods such as US have more complex channel models.
}
\begin{table*}[!t] \footnotesize
    \renewcommand{\arraystretch}{1.1}
    \caption{Comparison of different intra-body communication technologies. US refers to ultrasounds, IC to inductive coupling, CC to capacitive coupling and GC to galvanic coupling.}
    \centering
    \begin{tabular}{lcccc} 
    \hline
        \multirow{2}{*}{\textbf{Technology}}&\textbf{Operating}&\textbf{Max}&\textbf{Max}&\textbf{Data}\\
        &\textbf{Frequency}&\textbf{Attenuation}&\textbf{Distance}&\textbf{Rate}\\
        \hline
        \\
        US\textcolor{black}{~\cite{ref7,USrate,newUS1,newUS2,newUS3,newUS4,R1}} &1-100 MHz&100 \textcolor{black}{dB}~\cite{ref7}&\textcolor{black}{20 cm~\cite{R1}}&\textcolor{black}{Tens of Mbps}~\cite{USrate}\\
       \textcolor{black}{Microwave~\cite{FAT3,Yao-Hong2022IR-UWB}}&\textcolor{black}{2-9 GHz}&\textcolor{black}{25 dB~\cite{FAT3}}&10 cm~\cite{FAT3}& \textcolor{black}{up to Gbps~\cite{Yao-Hong2022IR-UWB}}\\ %p.58 e p. 84 di~\cite{FAT3}
        \textcolor{black}{mmWaves}~\cite{mmW2}&30-300 GHz&80 \textcolor{black}{dB}&50 cm&Gbps\\
        THz~\cite{Thz-Opto-LinkBudget}&0.1-10 THz %\textcolor{red}{(1 THz)}
        &66 \textcolor{black}{dB} & 1 mm&Tbps \\ %Sec. IV.E
        Optical~\cite{Thz-Opto-LinkBudget}&400-750 THz 
        %\textcolor{red}{(500 THz)}
        & 97 \textcolor{black}{dB}& 10 $\mu$m&Tbps \\
        IC~\cite{ICTable}&DC to 50 MHz&35 \textcolor{black}{dB}&130 cm & -\\
        CC~\cite{CCTab1,CCTab2,CCTab3}&100 kHz-120 MHz&65 \textcolor{black}{dB}~\cite{CCTab1}&170 cm~\cite{CCTab2}&\textcolor{black}{Tens of Mbps}~\cite{CCTab3}\\
        GC~\cite{GCTab1,GCTab2}&10 kHz-10 MHz&65 \textcolor{black}{dB}~\cite{GCTab1}& 15 cm~\cite{GCTab1}&\textcolor{black}{Mbps}~\cite{GCTab2}\\
        \hline
    \end{tabular}\label{TabTechs}
\end{table*}

\textcolor{black}{Table \ref{TabTechs} reports metrics derived from different sources, listing the maximum achieved values for each metric, which is an indication of link performance. These metrics play a crucial role in selecting the proper technology enabling the creation of an implantable network for specific applications. Based on application requirements such as latency, data rate, and required bit error rate (BER), the most appropriate intra-body communication method can be chosen. \textcolor{black}{All the above technologies indeed show different benefits and drawbacks,  depending on the specific addressed application scenario.} As shown in Table \ref{TabTechs}, high EM methods exhibit large attenuation leading to very short transmission distance. Hence, they would not be suitable to cover long distances in order to send data from implants to a wearable data collector for further processing in external IoMT networks. For this application, coupling technologies working at low EM frequencies would be preferable since they exhibit lower signal attenuation compared with high frequency solutions. At the same time, the very short wavelength of light, in the order of hundreds of nanometers, allows optical-based techniques operating at high EM frequencies to interact at the nanoscale with a single neuron. This would be of great importance for next-generation precise and localized neural stimulation, which is not yet possible with the current stimulation methods.}

\textcolor{black}{Fig.~\ref{fig:Fig_compare} summarizes the constraints for a transceiver design comparing all the technologies. As mentioned above, no technique outperforms  all the others. Each technique offers a unique set of trade-offs that influence specific application choices. For example, wearable use-cases requiring on-skin propagation of data and short transmission distances may be supported through multiple techniques. Anyhow, in case high data rate is a stringent requirement, US or high EM methods would be more successful since coupling methods operating at low EM frequencies offer lower data rate. From an interference and security perspective, US and GC offer the best performance. Indeed, their physical principles allow the signal to be confined inside the body so that it results to be interference-free from external signals and no malicious user can detect the data from the external environment. US and GC also show great safety features since they require very low transmission power and do not expose tissues to high temperature. The simpler transceiver design is guaranteed by GC due to the simplistic communication channel (see Table \ref{TabChannelTechs}), in line with the requirement of intra-body networks.
The next two sections report the most significant theoretical results and a subset of testbeds that have been already experimented in real settings.}

\section{Intra-Body Communications for Nervous System Applications}
\label{MacroApp}

\begin{figure}[!t]
\centering
	\includegraphics[width=.45\textwidth]{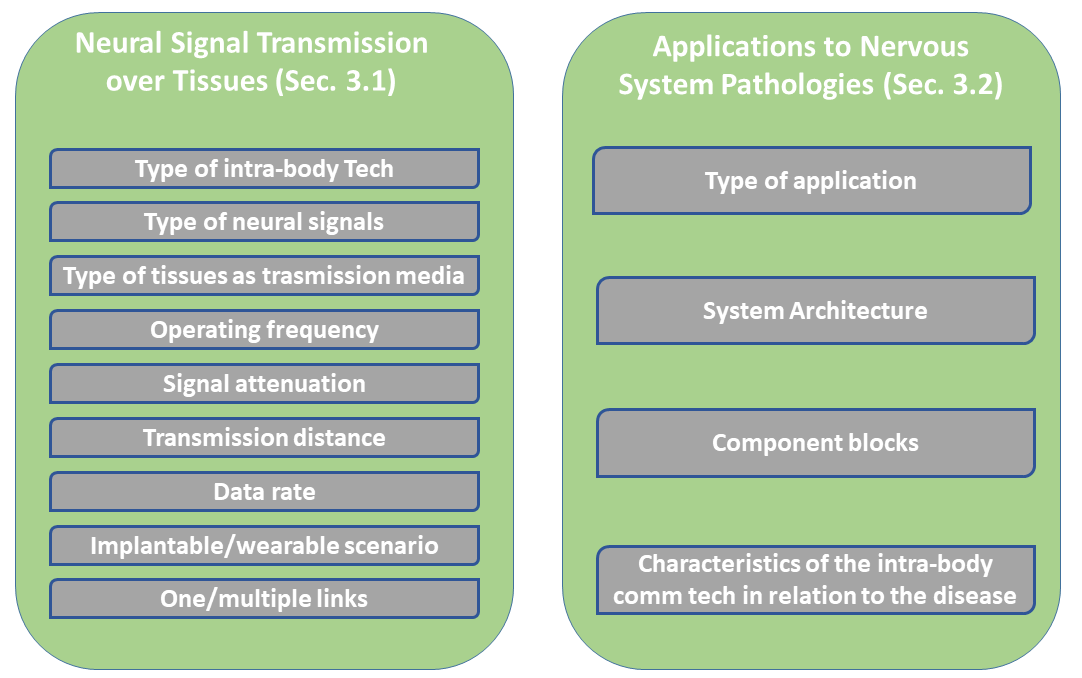}
	\caption{\textcolor{black}{Graphical representation of the topics covered in Sec.~\ref{MacroApp}: Intra-Body Communications for Nervous System Applications.}} \label{fig:sec3}
\end{figure}

\textcolor{black}{In this section we focus on applications of intra-body communication technologies to the nervous system. 
As shown in Fig.~\ref{fig:sec3}, Sec.~\ref{NeuroTX} focuses on neural signal transmissions, while Sec.~\ref{NeuroAPP} on solutions for targeting specific neural pathologies. Specifically, Sec.~\ref{NeuroTX} %analyze neural signal transmitted by means of intra-body technologies.
investigates the feasibility of employing intra-body technologies to transmit the physiological neural signals and recover them unaltered. In particular, we consider several elements, such as (i) the type of transmitted neural signals, e.g. electromyography signals acquired at muscle level or neural signals acquired in the brain, (ii) the body tissues over which the neural signals were sent, (iii) the operating frequency, (iv) the obtained signal attenuation, (v) the transmission distance, (vi) the achieved data rate, (vii) the communication configuration consisting in one or multiple links, and (viii) the type of scenario, i.e., fully implanted or wearable-implanted scenario. In the first case, implants transmitted data among them, while in the latter case the implants sent data towards a wearable device for data collection. 
%that, after acquisition, are transmitted by means of intra-body technologies over different body media. 
%The systems  being analyzed consist of one or multiple communication links. 
Sec.~\ref{NeuroAPP} instead focuses on neural pathologies. Here, we present a number of existing solutions describing: (i) the type of application, (ii) the communication system architecture, (iii) its component blocks, (iv) the specific characteristics of the communication technology chosen for targeting a particular disease.
%based on intra-body communication technologies that leverage on the features of the employed technology, as well as on the communication system architecture specifically designed for targeting a particular neural application.
Both Sec.~\ref{NeuroTX} and Sec.~\ref{NeuroAPP} are divided according to the intra-body technology used.}

\subsection{Neural Signal Transmission over Tissues}
\label{NeuroTX}
%il taglio che diamo è solo cose con intrabody comm tech per cui non mettiamo analisi spikes a macro level, accennare solamente!
%Differently from MC, macro-scale technologies consider the neural signal at a higher level, without dealing with the transport process of the involved molecules among neurons and their gap junctions. 

\textcolor{black}{Action potentials ('spikes') are the primary means of communication between neurons in the brain. Action potentials generate a transmembrane potential that can be detected by an electrical conductor in the extracellular medium near the neuron. Direct electrical coupling between sensor and neural tissue enables temporally precise recording of single-unit firing in conjunction with population synaptic activity. Neural probes, which convert extracellular ionic currents into electrical signals, are widely used in neural activity recording~\cite{li2023recent}.} The neuro-spike signals are usually characterized as a train of pulses with a peak-to-peak voltage of $100\,$mV, a pulse duration around $2\,$ms, and a rate of tens of Hz.

\textcolor{black}{Extraneural electrodes are typically characterized by a low signal-to-noise ratio (SNR). Therefore, acquired signals must be carefully pre-processed to filter out noise sources to select the frequencies of interest~\cite{coviello2022neural}.}
\textcolor{black}{A bio-inspired, event-driven neuromorphic sensing system (NSS) is proposed in \cite{Yao-Hong2022Sensing}. The NSS chip is able to perform on-chip feature extraction and “send-on-delta” pulse-based transmission, for peripheral nerve neural recording applications.}

%\subsection{Signal Transmission}
%\textbf{Transmission of Neural Signals:}
\textcolor{black}{In the following we present a number of studies on neural signals transmission that employed different intra-body communication technologies and considered} different types of \emph{tissues} as \emph{transmission media}, e.g., muscle, skin, fat, etc. 
\textcolor{black}{Also, the possibility to use the nervous tissues themselves, particularly the brain, to send data, has been investigated. Recorded neural signals were transmitted over one or multiple tissues. 
%TOLTO-MajorRev: in view of remote analysis and proper actuation outside the body through IoMT architectures. 
} 
%TOLTO-MajorRev: including exploitation of electromagnetic coupling and ultrasounds. 
%in view of implantable applications. 
%TOLTO-MajorRev: In the following section, we analyze a number of specific solutions for neural signal transmission over tissues employing variable technologies among those discussed previously. 
%%%%%%%
\subsubsection{Ultrasounds} 
\textcolor{black}{This technique was used to send neural recordings from miniaturized implants placed in deep brain regions to a transducer placed externally~\cite{Das2020,133Das2020}. 
%TOLTO-MajorRev: The studied configurations open the way to counteract several neural diseases, such as Alzheimer and Parkinson. 
Specifically, a pre-recorded neural signal, acquired from an awake-behaving rat motor cortex, was sent through in-vitro transmission with a data rate higher than $35\,$kb/s/implant. For in-vitro verification of a single communication link, an implant was placed at the depth of $45\,$mm in a tissue phantom and a wireless transmission of the recorded neural data was performed from the implant to the external interrogator. Around $0.5\,$dB/cm attenuation was obtained at $2\,$MHz~\cite{133Das2020}.} Preliminary evaluation of multiple links were also conducted by considering \textcolor{black}{two implanted transmitters} placed in a phantom at a depth of $50\,$mm with a $2\,$mm separation, transmitting single tones (414 and $313\,$Hz), at different operating frequencies ($55$ and $27.5\,$kHz)  to the same external receiver~\cite{133Das2020}. \textcolor{black}{These testings prove the feasibility of using US for \emph{intra-brain} communications.}

A similar type of neural dust solution was also experimentally validated in vivo in the rat peripheral system and skeletal muscle, reporting electroneurogram recordings from the sciatic nerve and electromyographic recordings from the gastrocnemius muscle~\cite{144Das2020}. 

The attenuation of US in brain tissue was also tested, proving that it is much smaller than the loss of electromagnetic waves from $3\,$MHz up to $430\,$THz~\cite{OPTO,Das2020}:
indeed it was observed that it is in the order of $1\,$dB at $2\,$mm for US operating at $10\,$MHz, as compared to $20\,$dB for EM waves at $10\,$GHz. 
\textcolor{black}{Observations on the reduction of peak acoustic pressure when passing across the skull were also done, showing the good US transcranial ability achievable \cite{newUS2}.}

\subsubsection{High EM Frequencies}  

Besides testing brain conductivity at frequencies up to hundred of THz~\cite{OPTO,Das2020}, neural recorded data are planned to be sent via microwaves inside the fat tissue~\cite{FAT2}.
At microwaves, random data transmission were evaluated for fat communication operating at 2 GHz~\cite{FAT3} in one link mode achieving $250\,$kb/s, %pag 84/118 di \cite{FAT3} 
and, recently, physiological data, i.e., recorded intracranial pressure that measures the health of a brain after an injury, were experimentally transmitted over the in vivo fat tissue of a pig. 
%\textcolor{black}{In \cite{Yao-Hong2022IR-UWB}, an impulse-radio ultra-wideband (IR-UWB) wireless telemetry system, for intracortical neural sensing interfaces, can reach a high data rate of 1.66 Gbps, in the 6–9-GHz UWB band.}
\textcolor{black}{In \cite{Yao-Hong2022IR-UWB} implantable impulse-radio ultra-wideband (IR-UWB) wireless systems, operating in a 6–9-GHz UWB band, are considered for intracortical neural sensing interfaces. These technologies allow a high energy efficiency of 5.8 pJ/bit, and a high data rate of 1.66 Gbps, obtained through 8-phase-shift keying (PSK) and 4-pulse-position modulation (PPM).}

At the higher optical frequencies, given the ability of these to interact with neural cells for the short wavelengths in the order of hundred nanometers~\cite{OPTO1}, the research is focusing on the capability to stimulate single neurons with light, rather than using neural data transmission within tissues. \textcolor{black}{More details about optical stimulation of neurons are given in  Sec.~\ref{NeuroAPP} and in Sec.~\ref{SecTestbeds}.}

\subsubsection{Low EM Frequencies} 

%\textcolor{red}{QUESTA SEZIONE RISPETTO ALLE ALTRE IN EFFETTI SEMBRA MOLTO LUNGA. FORSE SI POTREBBE UN POCHINO RIDURRE OPPURE CE NE FREGHIAMO E LA LASCIMAO COSI'}

At low frequencies, more studies were conducted for intra-body communication of neural data over heterogeneous multiple tissues, as well as on nervous tissues themselves.

GC was used to transmit neural and neuro-muscular signals, over heterogeneous tissues. 
In particular, data from epidural recording sites were sent to an external system through intra-skin communication at $10\,$kb/s with a channel attenuation of $17\,$dB~\cite{GC5}. The implant was placed inside a cavity in the bone and was completely covered by the skin to prevent any possible infection. The implant recorded neural activities from sixteen flexible epidural electrodes and transmitted them using skin as conductive medium~\cite{GC5}. Two concurrent transmissions through the skin were tested from two implants, $1\,$cm far apart from each other, to a single receiver.
%It is worth to note that the transmitted signals does not affect or interfere with neural activities, since neurons are transparent for high frequency signals ($>$ 100kHz)~\cite{GC5}.

%\textcolor{black}{In \cite{Yao-Hong2022GC} a new as a new high-speed trans-dural data transmission, based on GC, is presented The proposed in order to replace the tethered wires connected in between implants on the cortex and above the skull, allowing the brain implant to be “free-floating” for minimizing brain tissue damage.}
\textcolor{black}{A transmission method based on digital-impulse GC has been proposed for high-speed trans-dural data transmission (from cortex to the skull) \cite{Yao-Hong2022GC}. This technology allows the brain implant to be “free-floating” for minimizing brain tissue damage, showing that the trans-dural channel has a wide frequency response up to 250 MHz. In \cite{Yao-Hong2022Sensing}, a bio-inspired, event-driven neuromorphic sensing system (NSS) has been presented, considering an on-chip feature extraction and “send-on-delta” pulse-based transmission for peripheral nerve neural recording applications.}

In~\cite{GC7}, electromyography pre-recorded signals, i.e. neuro-muscular signals acquired at muscle level, were transmitted over muscle, fat and skin tissues obtaining few kb/s of rate employing GC. The operating frequency was set to $10\,$kHz and both implanted and wearable configurations were tested. The implanted scenario was performed on an ex-vivo chicken breast with two pairs of $0.5\,$mm electrodes (one for the transmitter and one for the receiver) implanted in the meat, while the wearable configuration employed $3\,$cm commercial electrodes placed on a leg skin. In the latter scenario the very low currents were flowing within skin, fat and muscle. Since the \textcolor{black}{electrode's size} influences the achievable communication range, the distances were tested up to 10 cm for the implanted scenario and $25\,$cm for the wearable configuration achieving almost \textcolor{black}{error-free} performance with the developed signal processing techniques~\cite{GC7}.

%\textbf{Transmission over Nervous Tissues:}
%(brain GC/CC (GC3) e image brain GC/CC (GC4))\\
Furthermore, the conductivity properties of the nervous tissues as communication media themselves were tested. In particular, the brain tissue has been evaluated to transmit data between neural implants and a system for data processing in a rat~\cite{GC3}.
Recorded neural data were successfully sent and recovered through \textcolor{black}{intra-brain communication} via an implantable version of CC. Two implanted transmitters separated by 15 mm, working at different frequencies in the order of hundred kHz, sent neural data  to an implanted receiver with an average BER around $10^{-5}$.
The goal was to evaluate the use of brain as conductive medium to send data without interfering with the natural neural signals. 
Artifacts or abnormal neural activities were not observed indicating, qualitatively, that intra-brain communication does not affect neural activities~\cite{GC3}.

Transmission tests of all the above technologies, over both nervous and non-nervous tissues, prove the feasibility to use them inside the body. Neverthless, these technologies have different features, with different values for the achievable distance, data rate, attenuation (see Table \ref{TabTechs}). \textcolor{black}{
%As will be underlined in Sec.~\ref{SecTestbeds} describing the existing testbeds, coupling transceivers operating at low EM frequencies are less complex than US and high EM methods. Coupling solutions also show lower power consumption covering longer distances due to lower signal attenuation within the tissues, while ultrasonic and optical approaches achieve higher data rates with the ability of targeting single neurons, at the cost of more complex transceivers. 
Hence, as will be underlined in Sec.~\ref{SecTestbeds}, depending on the target application with its specific requirements, a technology may be preferable as compared to the others.}

\begin{figure}[!t]
    \centering
	\includegraphics[width=0.4\textwidth]{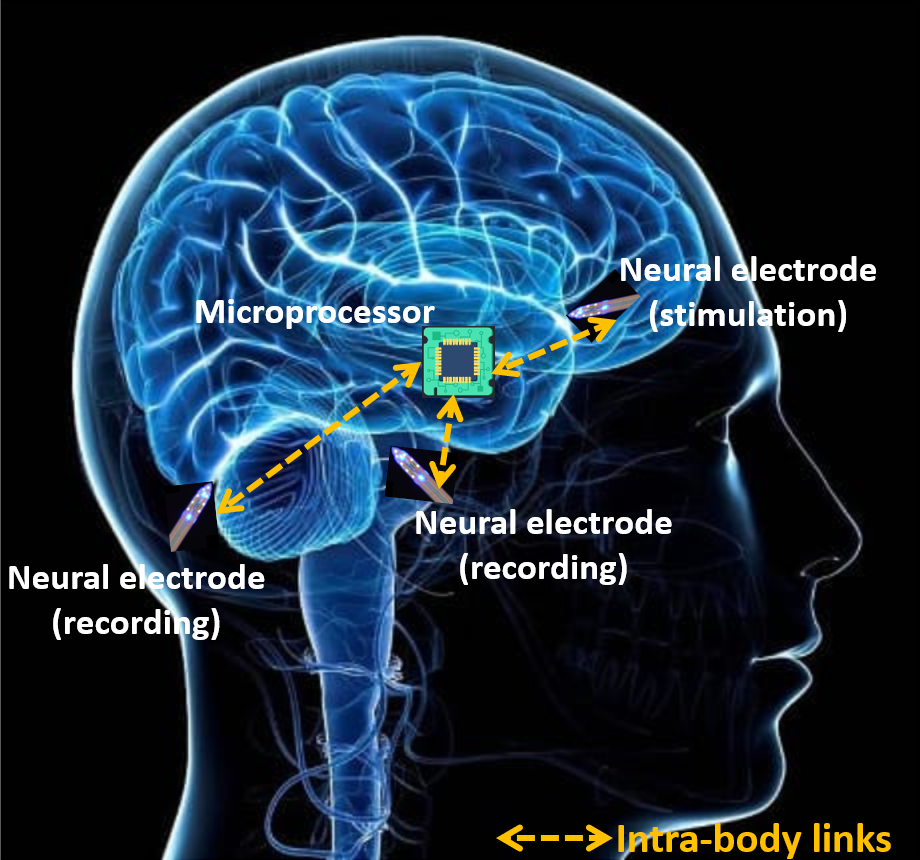}
	\caption{Conceptual scheme of a combined neural recording and stimulation that leverages on fully implantable devices and intra-body communication links.} \label{fig:NeuroApplication}
\end{figure}

\subsection{Applications to Nervous System Pathologies}
\label{NeuroAPP}

One of the most relevant applications of intra-body communication technologies is related to neural recording and stimuli. The detection of neural signals, often performed directly in the brain, is indeed of primary importance for both disease detection and  treatment. %and it is used to take the appropriate actions to stimulate the neurons by delivering neural pulses. 
Neural stimulation consists in pulse delivery and is widely used in neuroscience for  therapeutic purposes on patients with mental disorders, depression, Parkinson, and Alzheimer~\cite{OPTO}. Fig.~\ref{fig:NeuroApplication} depicts a possible application that integrates both neural recording and stimulation leveraging on implantable micro devices (neural opto electrodes and microprocessor) and intra-body communication links. 
\textcolor{black}{In the rest of this section we will detail the literature in the field, with a particular focus on %TOLTO-MajorRev: the use of non RF technologies, 
the architecture of the communication system, its component blocks, and the specific features of the intra-body technology chosen for targeting a particular disorder.}

\subsubsection{Ultrasounds} This technology was employed for neural recording at the brain level by combining power transfer and communication through brain-machine interfaces~\cite{Das2020}. %that can be useful to treat disorders such as epilepsy~\cite{Das2020}. 
The envisioned system architecture consists of several micro sensors, called {\it neural dusts}, that detect local extracellular electrophysiological information in different brain locations and send it to a subcranial interrogator. 
These use of small devices can minimize tissue damage and scar formation and, consequently, also neuro-inflammatory response. The implants achieve simultaneous power and data delivery with an external transducer~\cite{Das2020,133Das2020}.
This feature is fundamental since it allows to miniaturize wireless implants below millimeter scales~\cite{133Das2020}. Indeed, separate power and data communication links were proposed for ultrasonic implants~\cite{US15di133Das2020,US16di133Das2020}, which, however, need two ultrasonic resonators that limit the miniaturization~\cite{133Das2020}.  
A single ultrasonic link is instead used in~\cite{133Das2020} for both power and uplink data transmission using a pulse-echo scheme, thus avoiding the need for a secondary resonator.
An ultrasonic pulse is launched towards the implant that starts resonating and harversting energy. Shortly after that, the implants wake up and begin recording neural signals that are sent back by modulating the amplitude of the echo, that is travelling from the implant towards the interrogator.
Since a network of sub-mm implants is conceived that simultaneously powers up and performs data back-telemetry, for the purpose of uplink data transmission each implant has a unique orthogonal subcarrier that uses code-division multiplexing while modulating the amplitude of its echo~\cite{133Das2020}.

These sub-mm ultrasonic devices were also successfully employed to record neural data from peripheral nerves~\cite{144Das2020} and could also be used to stimulate nerves and muscles, which is useful to treat disorders such as epilepsy or to stimulate the immune system or tamp down inflammation~\cite{Das2020,US3,US4}. 

\textcolor{black}{A feasibility analysis on the use of US in the framework of peripheral nervous system stimulation has been provided in \cite{newUS1}. In this specific case, an analysis about the specific ultrasound parameters which result effective in stimulating excitable cells has been carried out. A single transducer operating in a variable range of frequencies has been employed showing that low frequencies around 300 kHz are more effective in stimulating tactile and nociceptive responses in human beings as compared to use of frequencies around 900 kHz or 1MHz. These effects were observed independently of the specific pressure being considered and independently of the mode of operation, being it continuous or pulsed. This distinction about the impact of different frequencies in terms of human perception drives the choice related to transducer hardware design when considering applications to the peripheral nervous system. In \cite{newUS2} also ultrasonic neuromodulation on large animals, like primates, has been tested using arrays of transducers to stimulate various brain areas.
A 2D plane array, consisting of tens of 256-element square modules was designed and fabricated and the central frequency of the system chosen around 1.04 MHz. Tests were executed in a water tank on an ex vivo macaque skull. 
}

\textcolor{black}{In \cite{newUS4} US technology is specifically applied in the context of other nervous systems-related pathologies, such as those associated with use of prosthetic hands. More in depth, the focus is on the prediction of wrist and hand movements.  A Gaussian process model and a multi-task deep learning algorithm are employed to simultaneously predict wrist rotation (pronation/supination) and finger gestures for transradial amputees via a wearable ultrasound array. A rich comparative analysis shows the superiority of US over surface electromyography showing the potential effectiveness of US in next generation prosthetic control. The US-based adaptive prosthetic control dataset (Ultra-Pro) is also in the process of being released to the scientific community for further contributing to studies in this field.}

\subsubsection{High EM Frequencies}
In the past years, several implants have been developed for neural recording and stimulation and some of them have been wirelessly interconnected with external devices through EM technologies~\cite{BioCyber2021,65diBioCyber2021,66diBioCyber2021}. 
%However, these works mainly focus on the specific design of the neural interface rather than considering an intra-body communication point of view. Specifically,  wireless technology interfacing is considered only as an instrumental last step  to power the implant and transfer data outside.
%\textcolor{red}{qui c' una parte commentata a seguire che si potrebbe usare. comunque forse non sarei così drastcia nel commento e li citerei.}
These works focus on the design of the neural interface and use a wireless technology interfacing to power the implant and transfer data outside.
%\textcolor{red}{ULTIMO PARAGRAFO SEC. 4.1 DI \cite{BioCyber2021} (Review-Arxiv-Bio-CyberInterface-Sasi.pdf), check freq sia qui in app che testbed?! \cite{65diBioCyber2021,66diBioCyber2021}:
%The brain machine interface can be wirelessly interconnected with external devices. For instance, a chip-less wireless neural probe system was proposed for both performing stimulation and reading of neural activities simultaneously [65]. This device consist of two antennas, which allows to transfer data and wirelessly power the system, and have the following dimensions: 7 mm, 50 $\mu$m, and 31 $\mu$m (length, width, and thickness, respectively).This device have a maximum peak stimulation voltage that can reach to 100 $\mu$m. Moreover, the device does not sense any chemicals, but detect neuron’s electrical pulses. A similar approach was considered to design a microsystem based on electrocorticography and placed it on the surface of the cerebral cortex [66]. This microsystem consists of a 64-channel electrode array and a flexible antenna, and has size of 2:4mm $\times$ 2:4 mm.The device does not influence on the cell, but performs data acquisition, wireless power and data transmission. This device is suitable for long term monitoring due to the use of biocompatible materials.
Specifically, the brain machine interface was wirelessly interconnected with external devices. 
A chip-less wireless neural probe system was developed to perform simultaneous recording and stimulation of neural activities~\cite{65diBioCyber2021}. 
This device consists of two antennas, to transfer data and wirelessly power the system, and does not acquires any chemicals, but senses neuron’s
electrical pulses~\cite{BioCyber2021,65diBioCyber2021}. 
A similar method is used in ~\cite{66diBioCyber2021} to design a microsystem based on electrocorticography, consisting
of a 64-channel electrode array and a flexible antenna.
The device does not affect the cell, while performing data acquisition, wireless power and data transmission. This device is suitable for long term monitoring since bio-compatible materials are used.

Recently, microwaves were proposed to be used for the transmission of sensed neural data inside the fat layer of the body, with the final goal of developing a battery-free high-speed wireless in-body communication platform for brain-machine-body connectivity~\cite{FAT2}. 
While in the past, implanted medical devices were connected to the Internet,
the B-CRATOS project aims at developing systems that will communicate
without involving an external computer~\cite{FAT2}.
As an application example, the aim is to create a direct connection between
a person’s artificial arm and his/her brain so that it can sense its presence and experience control over this body part. This requires an implant in the brain and an intra-body communication link between the artificial arm and the implant in the brain that is established with devices working at microwaves around 5.8 GHz over distances of approximately one meter~\cite{FAT2}. The project is at an early stage but several applications are envisioned, such as neuroprosthetics to bypass damaged circuits for the restoration of missing biological functions, or electroceuticals that are new therapeutic agents able to deliver neural impulses to the neural circuits of organs, to be used in place of drugs.
%\textcolor{red}{electroceuticals in order to organ function modulation through neural circuits instead of drugs, and brain-computer interfaces for brain plasticity through machine learning. (ANNA: spiegare meglio)}

Furthermore, a microwave beamforming for non-invasive brain stimulation was proposed~\cite{MicrowavesBeamBrain}. In general, implanted electrodes configuration or wearable solution may be used to stimulate the nerve fiber of specific regions (cerebral cortex) by electric or electromagnetic fields, whose main problems are focusing and penetration. \textcolor{black}{Although recently optical and ultrasonic methods were gained attention for such application,} they show some limits: optical solutions are highly accurate but suffer from a low penetration depth, and US technology show a good penetration depth but poor focusing capability.
In~\cite{MicrowavesBeamBrain}, an array structure antenna based on microwaves, which is placed externally to the body, is used to transmit and focus EM power in the chosen area, with good focusing and penetration depth.
A method was developed based on pulse shaping with space-frequency decomposition of the time reversal operator method (DORT) to select proper amplitude and phase for each location.
Simulations were performed with a 3D brain phantom mode showing satisfactory performance. 
The target point was selected at 10 cm below the skull, and the array antenna was construct in the cylindrical structure with 6$\times$4 dipole array with 5 cm spacing around the skull.
The large array structure allowed to improve the focusing.
The frequency of stimulation was set to $1\,$GHz to ensure an appropriate penetration depth.

Lately, some solutions employed optical EM frequencies for deep brain stimulation, with a particular focus on optogenetic,s as a better option to stimulate neurons with light compared to conventional electrical stimulation~\cite{OPTO}. 
%Since here the focus is on the medical application, in the following we summarize together a number of solutions designed for the same application (i.e. optogenetic brain stimulation), without going in the details of each of them \textcolor{red}{(siete d'accordo?)}.
In particular, while existing solutions considered the insertion of wired optical cables into the skull, some new trends propose a miniaturized device equipped with an LED designed to wirelessly stimulate the genetically engineered neurons, which are sensitive to light at a particular wavelength~\cite{OPTO,3diOPTO,4diOPTO,5diOPTO,Montgomery2015,Park2015}. 
Although with some differences due to the chosen technology and the developed communication protocol, the general system architecture for optogenetic solutions consists of an external base station for energy powering and control to activate the implanted LED, which communicates to neurons through light.
The developed solutions include head mounted unit or fully implantable units embedded into the brain or nervous system, and may employ different wireless EM communication technologies, including infrared (IR), high frequency/near field communication (HF/NFC), and ultra high frequency (UHF). The chosen wireless technology reflects in different size of the device, propagation characteristics in the medium, and power sufficiency.
As the frequency goes up, the antenna size becomes smaller, hence HF and UHF technologies result to be more appealing for device miniaturization~\cite{OPTO}.

Briefly, IR optogenetics solutions ($300\,$GHz–$430\,$THz) have the advantage of low power consumption and multi-band transmissions at the cost of the need of line of sight (LoS) between the base station and the implanted unit and the requirement of a battery unit for the head unit~\cite{3diOPTO,4diOPTO}.
HF methods (3–30$\,$MHz) show good properties of medium propagation loss in biological tissue, are cheap to manufacture and support energy harvesting circuitry. However, they require coil large dimension of around $1\,$cm and surface mounted chip (NFC)~\cite{5diOPTO}.
UHF solutions ($300\,$MHz–$3\,$GHz) allow smaller coil diameter than HF circuitry, are easy to manufacture and support energy harvesting circuitry. However, they experience high propagation loss in biological tissue~\cite{Montgomery2015,Park2015}.
As it will be detailed in Sec.~\ref{CombTechs}, an interesting alternative is the integration of optogenetics with US for energy powering and command control, since US experience lower attenuation in biological tissues at the cost of manufacturing complexity~\cite{OPTO,OPTO3}. 

The use of optogenetics can lead to precise single neuron stimulation that can lead to the proliferation of optogenetic micro-scale approaches to better treat neural diseases requiring neuronal stimulation.
%An optogenomics based solution is proposed, leveraging on the control of the genome function through light. Specifically, 
Finally, an emerging research area is optogenomics, which goes beyond neurons stimulation since it leverages on the control of the genome function through light.
Optogenomics allows to control organ and in particular brain development and functions, which would be achieved by developing novel directional light emitting nano-devices~\cite{OPTO4}. 

\subsubsection{Low EM Frequencies}
At these frequencies, some solutions were designed to overcome compromised peripheral nerves as well as to perform wireless brain stimulation and monitoring.

A GC based architecture to reactivate functions lost due to nerve compressions was developed in~\cite{GC7}. The goal was to overcome the neural signal interruption due to a nerve compression preventing muscles movements. 
Since different muscles are involved in a contraction, a pre-recorded signal acquired from an activated healthy muscle was transmitted via GC with the final purpose of triggering the muscle suffering nerve compression that was unable to receive its natural signal.
Here a receiver device may process the signal and use it as input to trigger the muscle. To test the feasibility of the proposed approach, pre-recorded electromyography data were sent in a person leg and successfully recovered at the receiver side~\cite{GC7}. 
The electromyography signals were acquired by means of needle electrodes inserted in the muscle of a person forearm during muscle contractions. 
The signal was recorded for 10 s with a sampling
frequency of 20 kHz, then 1000 mean values were calculated with a consequent net sampling rate of 100 Hz.
The electromyography signal was modulated with GC technology working at a frequency carrier of 10 kHz and detected at the receiver with almost \textcolor{black}{error-free} performance.

In~\cite{GC1} a system was envisioned based on GC intra-body communication that used a combination of highly specific electrodes, communicating directly with the nerves in the brain and with a relay node that collects data to be delivered outside the body.
The relay node may be implantable or placed on the surface of the body.
The approach is minimally invasive since the implanted electrodes do not need frequent replacement and low power consumption due to the use of galvanic coupled communication systems is needed. Also, exploitation of distributed
beam-forming approaches such as those developed in~\cite{GC8}, will allow use of an array of electrodes to design a beamforming solution  to increase the resolution of the data transmitted to and from the relay node to the nerves, as well as to target remote areas of the brain tissues without direct electrode contact~\cite{GC1}.  

%TESTO DA \cite{ReviewIBC2022} Sec. 4.2.3:
Since most neurological diseases affect several brain regions, the capability of monitoring neural activity and evaluate intra-region communication is essential for our understanding of dysfunctions~\cite{ReviewIBC2022}, to treat illness according to a more appropriate and personalized approach.
\begin{figure*}[!t]
\begin{center}
	\includegraphics[width=0.9\textwidth]{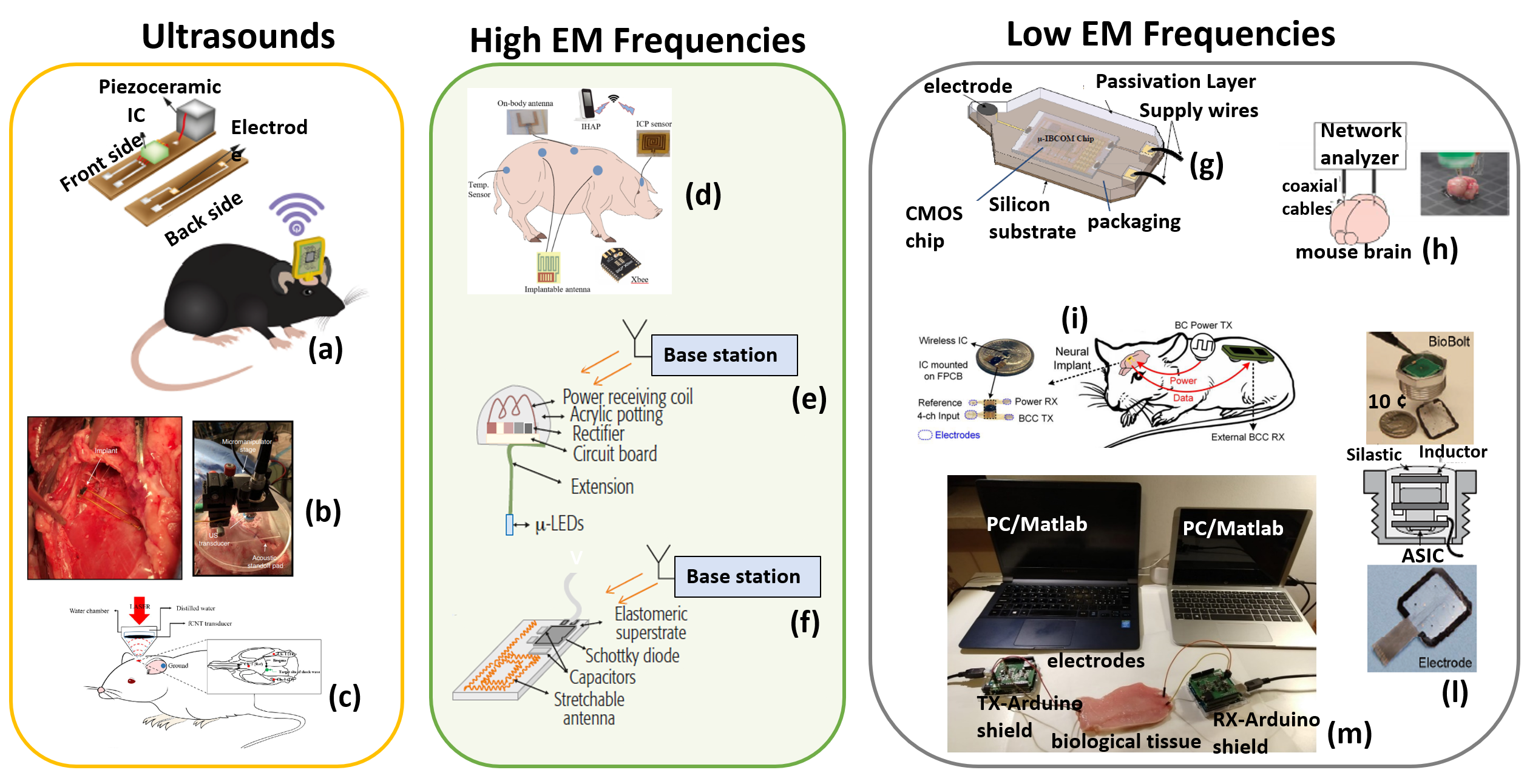}
	\caption{\textcolor{black}{Existing testbeds based on different intra-body communication technologies for nervous system applications: US (a)~\cite{133Das2020}, (b)~\cite{Mahara1}, (c) \cite{newUS3}; high EM frequencies (d)~\cite{ExpMicrowaves}, (e)~\cite{Montgomery2015,OPTO}, (f)~\cite{Park2015,OPTO}; low EM frequencies (g)~\cite{GC3}, (h)~\cite{GC4}, (i)~\cite{Lee}, (l)~\cite{GC5}, (m)~\cite{GC7}.}} \label{fig:TechTestbeds}
\end{center}
 \end{figure*}

% {\bf a seguire togliere perchè non abbiamo alcuna reference}
% \textcolor{black}{Finally, we propose an implantable neural communication system able to restore a facial paralysis due to a nerve's lesion, as shown in Fig.~\ref{fig:FacialPalsy}. 
% The system  design  leverages  on  combined ultrasound and galvanic coupling communication technologies to forward the signal detected from the  healthy  nerve  to  the injured contralater one in order to restore symmetric facial movements.}
% \begin{figure}
% \begin{center}
% 	\includegraphics[width=.9\columnwidth]{FacialPalsySoA}
% 	\caption{System architecture to overcome a facial palsy \textcolor{red}{(cambiare faccio e/o sfumarla)}} %\textcolor{red}{(Source: https://www.pacificneuroscienceinstitute.org/blog/facial-pain/7-questions-about-bells-palsy/)}}
% 	\label{fig:FacialPalsy}
% 	\end{center}
% \end{figure}

\section{Existing Neural Intra-Body Communication Testbeds}
\label{SecTestbeds}

%\textcolor{red}{(MAURIZIO: se opportuno, descrivere i riferimenti che avevi proposto MA SE SONO SIMULATORI ALLORA NON LI METTIAMO PERCHE' STIAMO METTOENDO SOLO EXPERIMENTAL TESTBEDS!!!:\\
%- Comsol Constadinou\\
%- PyPNS: Multiscale Simulation of a Peripheral Nerve in Python)\\}\\}

\textcolor{black}{In this section we discuss some testbeds that have been developed and tested with specific applications to the treatment of neural diseases.
Fig.~\ref{fig:TechTestbeds} summarizes the main features of the considered testbeds.} %which employ different types of signals, namely US, high frequency EM waves, and low frequency EM waves. In the rest of this section, these testbeds will be detailed.

\subsection{Ultrasounds}
%\textcolor{red}{Das2020 Sec. IV.F:}
\textcolor{black}{In the recent past, some studies have appeared on the use of US-based wireless neural implants.} Specifically, the Neural Dust~\cite{144Das2020} and Stim Dust~\cite{145Das2020} have been presented.  In line with this approach, in~\cite{133Das2020}, a neural dust platform was developed for neural recording based on low power CMOS circuitry and ultrasonic communication and powering (Fig.~\ref{fig:TechTestbeds}(a))
~\cite{133Das2020}. 
In particular, a sub-millimiter neural implant ($0.8\,$mm$^{3}$) was implemented with a tiny recording integrated circuit of $0.25\,$mm$^{2}$ size. A single piezoceramic resonator was used for both ultrasonic powering and data transmission, which allowed to reduce the miniaturization of the device.
The implant could operate at a depth of $5\,$cm, allowing neural recording
from deep brain regions~\cite{133Das2020} and most peripheral nerves~\cite{144Das2020}. The implants achieved simultaneous powering and data delivery with a cheap unfocused single-element transducer placed externally.
Thanks to the limited size of the device, it is indeed possible to improve the spatio-temporal resolution in case of
distributed recording with multiple implants.
The developed device has low power consumption since its integrated circuit dissipates only $37.7\,\mu$W and neural recording front end consumes $4\,\mu$W~\cite{133Das2020}. 

\textcolor{black}{Recently, in~\cite{Mahara1} a system reporting was presented, characterized by minimal invasiveness. The system monitored deep-tissue $O_2$ real time data from centimeter scale depths in sheep.}
Indeed, even a slight reduction in oxygen supply to the brain, so-called hypoxia, impairs behavioral responses and alertness levels.
\textcolor{black}{This system consisted of a  $4.5\,$mm$^3$, wireless, ultrasound-powered implantable luminescence $O_2$ sensor 
%incorporating a luminescence sensor $\mu$-LED sensor embedded with a single piezoelectric ceramic transducer, combined with 
and an external transceiver (Fig.~\ref{fig:TechTestbeds}(b))~\cite{Mahara1}.
The system used an $O_2$ sensor implant that incorporated a single piezoelectric ceramic and a luminescence sensor consisting of a \textcolor{black}{$\mu$-LED},
an O2-sensing film, an optical filter and a state-of-the-art integrated
circuit~\cite{Mahara1}.
The external transceiver was needed for both wireless power and bidirectional data transfer.} This device was thought for application in terms of  bidirectional data transfer for deep-tissue oxygenation monitoring, also thanks to the small size (weight of approximately 17.4 mg) potentially leading to minimal tissue damage in case of implant. Both in vivo testing on sheeps as well as ex vivo testing on porcine tissues were carried out.
\textcolor{black}{Another interesting example of the application of US to neuromodulation was proposed in the testbed described in \cite{newUS3} (Fig.~\ref{fig:TechTestbeds}(c)). Indeed, focused US can deliver acoustic energy to local regions of the brain, including deep brain. Also, by tuning appropriate parameters of the beam, it is possible to activate or inhibit nerves. A typical application of this principle is precise therapeutic US delivery. Specifically, shock wave execution was implemented with a tight focal spot for limited acoustic exposure of small areas for applications to precise targeting \cite{newUS3}. The interesting feature of this work is also associated with the use of focused Carbon Nanotube (fCNT) transducers and the measurement of electroencephalographic (EEG) signals in rat brains.
A laser was applied to the fCNT transducer %(F# = 0.7, focal length = 1.4 cm)
which was immersed in water, so that the fCNT layer produced a laser-generated focused ultrasound (LGFUS). A pulse laser system (Tribeam, Jeisys, Medical Inc., Seoul, Korea) with a wavelength of 532 nm and energy of 350 mJ was used to generate a shock pulse. %with a PRF of 5 Hz. 
The shock wave, generated by the carbon nanotube-polydimethylsiloxane (CNT-PDMS) composite transducer, had a beam width of 2.5 mm in the axial direction and 0.70 mm in the lateral direction \cite{newUS3}.
}

\subsection{High EM Frequencies}
At microwaves, for the first time, an end-to-end
transmission of physiological data was experimentally tested from implanted antennas mimicking sensors to a cloud-enabled aggregator device using
fat-microwave communication~\cite{ExpMicrowaves} (Fig.~\ref{fig:TechTestbeds}(d)). The network consisted of two implanted nodes and one node on the body. The implanted devices were made of antennas placed in the fat tissue of a live porcine that transmit physiological data pre-recorded from two sensors.
\textcolor{black}{The on-body device encrypted the sensor data and sent it via Bluetooth Low Energy to an Intel Health Application Platform (IHAP) device, which in turn forwarded the encrypted data to a web server~\cite{ExpMicrowaves}.}
The two sensors recorded intracranial pressure, an important metric to be measured  after a neural damage to make diagnosis of possible brain diseases. 
\textcolor{black}{The implantable antenna was designed for communication at 2.45 GHz, with Taconic's low-loss material used as substrate and Rogers RT Duroid as superstrate. The antenna was very low profile and lightweight~\cite{ExpMicrowaves}. 
The on-body antenna with flexible high-dielectric substrate was designed in Ansys for fat intra-body communication.}
The two implanted antennas were placed around 400 mm apart along the back area of the pig, together with the intracranial pressure sensor implanted on the cranium, and the aggregator node was placed on the skin in between the implanted nodes.
All the communication steps of the end-to-end system were implemented to communicate with an external server.
\textcolor{black}{A laptop PC running Ubuntu Linux 20.04 LTS was used to collect data and to operate various devices when needed during the experiment~\cite{ExpMicrowaves}.
Digi Xbee 3 transceiver units with U.FL antenna connectors were employed for wireless communication. They were configured to use an IEEE 802.15.4-compliant mode using the software XCTU, provided by the manufacturer Digi~\cite{ExpMicrowaves}.
The IHAP was used as an application software platform for healthcare, whose accompanying hardware is the Flex IoT Compute Engine. 
}
The experiments measured received signal strength indicator (RSSI) values of at most -$70\,$dBm at the aggregator on-body node that was operating at -$5\,$dBm output power. Tests revealed that the matching of the antennas with the fat channel could be improved to achieve a higher RSSI and to avoid unnecessary signal leakage~\cite{ExpMicrowaves}.

At optical frequencies, fully implantable solutions were developed, not requiring any external head mounted unit to operate.
Two implantable wireless optogenetic devices working at ultra high frequencies were implemented for neural stimulation, with a power-harvesting unit to receive the energy from an external RF base station located outside the body for device activation (Figs.~\ref{fig:TechTestbeds}(e) and~\ref{fig:TechTestbeds}(f))~\cite{Montgomery2015,Park2015}. 

A tiny device, smaller than the previous version of wireless optogenetic implants, 
was realized with a \textcolor{black}{weight} of $20$$\,–\,$$50\,$mg and a size of $10$$\,–\,$$25\,$mm$^{3}$ for brain, spinal cord, and peripheral circuits stimulation in mice (Fig.~\ref{fig:TechTestbeds}(e))~\cite{Montgomery2015,OPTO}. 
\textcolor{black}{The implant had an on-board circuit, including a three-turn coil to extract power, that drove a blue $\mu$-LED for neural stimulation. Acrylic encapsulation of the implant was used to resist to the biological degradation and electrically and to insulate the circuits~\cite{Montgomery2015}.}
%\textcolor{red}{(ref 7 di opto non scaricabile)}
The used $\mu$-LED was able to emit the light power density required for optogenetics excitation ($1$$\,–\,$$20\,$mW/mm$^2$), and had an optimum efficiency (emitted light power/input power) of 19\%. 
\textcolor{black}{The system required a large aluminum resonant cavity, which radiated RF frequency to both transmit power and control the implant. The size, geometry and resonant frequencies of the implant and cavity were optimized for mice, not for use in larger animals.
Anyway, it can be employed only in a controlled lab environment, not yet suitable for daily use in patients.}

\textcolor{black}{Another miniaturized optoelectronic system utilized stretchable filaments and a flexible polymer encapsulation,  which allow the implantation of the devices as soft injectable filaments, without the need for skeletal fixation. This allows experiments in regions where it would be impossible
to operate with other approaches. 
An implant was embedded and tested both into the spinal cord and peripheral nervous system of a mice (Fig.~\ref{fig:TechTestbeds}(f))~\cite{Park2015,OPTO}.}
The device consisted of an RF power-harvesting unit, a rectifier, a voltage multiplier, and a cellular-scale $470\,$nm LED, which communicates to the neuron with an optical power density of $10\,$mW/mm$^2$. 
\textcolor{black}{The antenna and LEDs were connected with serpentine Ti/Au electrical interconnects, and the circuit was encapsulated by polyimide and a low-modulus silicone elastomer~\cite{Park2015}.
The RF external base station might distribute around $2\,$W
for multiple-device activation within $20\,$cm range, in case of simultaneous operation of devices implanted into multiple animals.} 
However, these optical solutions are still on the mm scale and further effort is required to reach sub-mm scale for long term implantation~\cite{OPTO}.

\subsection{Low EM Frequencies}

%\textcolor{red}{@ANNA: AL SOLITO QUI SE VOGLIAMO EQUILIBRARE L'ESTENSIONE SI POTREBBE RIDURRE UN PO' PERCHE' E' MOLTO DETTAGLIATO, MA VEDI TU..}

CC and GC testbeds working at low EM frequencies were developed for brain, epidureal, and electromyographic data transmission. 

Intra-brain communication was established in~\cite{GC3}, which consists in a wireless signal transmission method that uses the brain itself as a conductive medium to transmit information. The objective was to send data and commands over the brain between neural implants and a system for data processing outside the brain. Preliminary studies were conducted between two micro transmitters and a receiver located in a rat brain~\cite{GC3}. Specifically, two CMOS chips were designed and fabricated to transmit prerecorded neural signals modulated at different frequencies to a receiver located in the rat brain, with a transmitter power consumption less than $65\,\mu$W and an active part of the chip %400\times270 $\mu$m$^2$ 
$0.1\,$mm$^2$ in size.
\textcolor{black}{The scheme of the prototype device is shown in Fig.~\ref{fig:TechTestbeds}(g)~\cite{GC3}.
The chip was packaged in a silicon substrate and consisted of one signal transmission electrode and two power supply wires
~\cite{GC3}.
The prototype chip was designed and fabricated using 0.25-$\mu$m
CMOS technology.}

\textcolor{black}{Similar to the configuration in~\cite{GC3}, two electrodes were inserted in a mouse brain, one as transmitter and another one as receiver, and a wireless image data transmission was demonstrated through the brain \cite{GC4}. The overall architecture consisted in a distributed implantable image sensor system, where several implantable sensors send the acquired imaging data to an extracorporeal device that forwards the data to the receiver system. However, the experimental procedure tested a single link transmission between an image sensor transmitter and a receiver.
A tiny CMOS image sensor was designed and fabricated using a 0.35 $\mu$m 2-poly 4-metal standard CMOS process of Austria Micro Systems. 
The sensor was based on a 3-transistor active pixel sensor.
The acquired signal was transmitted into the brain with a platinum electrode of
diameter 30 $\mu$m at the end of a coaxial cable. The signal was received with an identical electrode and recorded by an oscilloscope (Tektronics, DPO4034).  The received signal was then processed by a PC.
Since the ground of the transmitter image sensor should be separated from that of the receiver, baluns were inserted between the electrodes and the measurement equipment.
Going into more details, an image data from an implantable image sensor was sent via amplitude modulation at $50\,$MHz over the brain, %and the received signal was successfully recovered thus reconstructing the image.
with an inter-electrode separation of $8\,$mm between the transmitter and receiver electrode (Fig.~\ref{fig:TechTestbeds}(h)).} 
%To reduce electromagnetic coupling between the signal transmission paths, coaxial cables were employed.
%%The modulated signal was launched into the brain with a platinum electrode of 30 $\mu$m diameter, and 
%The signal was 
%%received with an identical electrode. %and recorded by an oscilloscope. 
%The spacing between the electrodes was around 8 mm.
%The received signal was demodulated by a PC and the image from the sensor was displayed on the screen.
The obtained SNR of the demodulated signal was $12\,$dB with an input power of -$20\,$dBm. To calculate SNR, the signal level was estimated from the difference between bright and dark pixels, while the noise level was obtained
from a dark image. Results confirmed that low power image transmission is feasible using the efficient propagation properties of living tissue, a fundamental step to develop distributed image sensor system for deep brain imaging as future application.

\textcolor{black}{In~\cite{Lee} a wireless neural implant was proposed that used body coupling for both data transfer and powering and incorporated a precision front-end for high-quality neural recording (Fig.~\ref{fig:TechTestbeds}(i))~\cite{Lee}.
%without resorting to skull positioning of the device and exploiting soft tissues propagation (Fig.~\ref{fig:TechTestbeds}(h))~\cite{Lee}.
%In this way very small electrodes are needed and compatibility of the implant with numerous different types of electrodes is possible. 
The implant only required small electrodes for data transmission and power delivery, and external devices with patch electrodes were placed far away from the implant exploiting body coupling to communicate with the implant. Specifically, the device for transferring energy to the implant and the receiver were placed in the back of the rat, which facilitates stable chronic in vivo recordings in freely behaving animals. 
The functioning of the wireless neural implant was validated through in vivo experiments on rats. The implant integrated circuit mounted on a flexible printed circuit board was covered with a biocompatible polymer and placed on the skull where the electrodes were attached subdurally to the surface of the cortex. A small rechargeable microbattery (MS621FE) was included and charged by an external device. 
The neural implant was fabricated in a 0.11 $\mu$m CMOS with a high-density capacitor option occupying a chip area of 4 $mm^2$. The receiver integrated circuit mounted on the rat's back was fabricated in a 0.18 $\mu$m CMOS. 
One of the main advantages of this system is related to the lack of perfect alignment between devices, as other types of solutions such as IC, and still ensures data/power delivery.}  The prototype provided satisfactory  wireless power delivery of 644$\,\mu$W and a data rate of up to $20.48\,$Mb/s at an output frequency of $40.96\,$MHz. From the energy perspective, the system supported a consumption of $32\,$pJ/b. In vivo experiments validated the effectiveness of the design.

\textcolor{black}{A minimally-invasive neural interface, BioBolt, for epidural recording was developed.
The BioBolt application specific integrated circuit was fabricated
using $0.25\,\mu$m CMOS technology with a low-power wireless data transmission using intra-skin communication from recording sites to an external system~\cite{GC5}. 
The entire electrical components were enclosed in a bolt-shaped titanium fixture. 
For electrical insulation, BioBolt was coated with parylene and filled with silastic.}
Two devices, each with a core area of $3\,$mm$^2$,
were designed to simultaneously send data to a receiver placed $10\,$cm far apart from them using intra-skin communication. The system consumes $160\,\mu$W for the communication driver and globally $365\,\mu$W (Fig.~\ref{fig:TechTestbeds}(l))
~\cite{GC5}.

Finally, a testbed was implemented to send neural data recorded at muscle level (electromyography signals) via GC, employing $0.5\,$mm electrodes implanted within an ex-vivo muscle tissue and a low transmission power in the order of~$\mu$W (Fig.~\ref{fig:TechTestbeds}(m))
~\cite{GC7}. The employed GC testbed leveraged on PC sound card since GC frequency range ($1\,$kHz - $100\,$MHz) includes the signals’ frequencies supported by the sound cards. Two PCs were employed to run the transmitter and receiver Matlab software, respectively.
\textcolor{black}{A Conexant's CX20723 sound card was employed for the transmitter and a Realtek ALC888S 7.1 Channel HD Audio peripheral board for the receiver. Both sound cards supported up to 24 bit and 192 KHz sampling frequency.}
The pre-recorded signals was modulated with a frequency carrier of 10 kHz and sent out from the PC to the biological ex-vivo tissue through the LINE OUT jack, which was connected to the electrodes to inject currents inside the tissue, and similarly at the receiver through the LINE IN jack.
Physical methods were developed in the transmitter/receiver, including frequency, phase, and time recovery.
Almost \textcolor{black}{error-free} performance were achieved varying the distance between transmitter and receiver up to 11 cm.
The normalized mean squared error between the electromyography signal to be transmitted and the reconstructed one at the receiver, was around $10^{-5}$.

\begin{table*}[!t] \footnotesize
    \renewcommand{\arraystretch}{1.1}
    \caption{Comparison of different intra-body communication experimental solutions for nervous system applications. US refers to ultrasounds, UHF to ultra high frequency, CC to capacitive coupling and GC to galvanic coupling.}

    \centering
    \begin{tabular}{llllll} 
    \hline
        \textbf{Technology}&\textbf{Frequency}&\textbf{Power}&\textbf{Data Rate}&\textbf{Application}\\
        \hline
        \\
        US~\cite{133Das2020} & $1.78$ MHz&$4\,\mu$W (recording),&$35\,$kb/s&Recording/\\
        && $37.7\,\mu$W (tot)&&BMI\\
        
        US~\cite{Mahara1}& 2 MHz & $<$$150$ $\mu$W(consumption/ & kb/s & Monitoring\\
        &&sensing$\&$trx)&&\\
        
        Microwaves~\cite{ExpMicrowaves}&$2.45$ GHz&$316\,\mu$W &$115200$ &Monitoring\\ %-5dBm=316 uW - 460 kbps sono 115200 baud se 16QAM per standard 802.15.4, con 16 stati di modulaz ) per symbol ho 115200*log2 16 - forse non usano 16QAM ma modulazione binaria per cui 115200x2=230kbps - per sicurezza lasciamo BAUD
        &&(output tx power)&symb/s&\\
        
        UHF~\cite{OPTO,Montgomery2015} & $300$ MHz-$3$ GHz &$1$-$20\,$mW/mm$^2$ &-&Neural \\
        &&(emitted density)&&stimuli\\
        
        UHF~\cite{OPTO,Park2015} & $300$ MHz-$3$ GHz & $10$ mW/mm$^2$&-&Neural \\
        &&(emitted density)&&stimuli\\
        
        CC~\cite{GC3} & $100$-$400$ kHz&$<10$$\,\mu$W (tx), & 100 kb/s&Neural \\
        &&$65$$\,\mu$W (trx)&&data tx \\
        
        CC~\cite{GC4}&50 MHz&10$\mu$W  %\textcolor{red}{$-20$ dBm (input power)}
        &min. 200 kb/s& Monitoring\\ % 32 frames/s con 60x60 pixels in 1 image, hp 8 bit per frame->60x60x8x32=921.6 kbps--metto 200KHz di clock ma modulano in analogico AM, non digitale--per sicurezza mettiamo min. 200 kbps
        %Image TX\\
        &&(input power)&&\\
        
        CC~\cite{Lee}&20.46 MHz&$8.6$ $\mu$W (recording),&640 kbps&Recording\\%16 bit per 4 canali, mandati serialebspedendo tutto con 10KHz freq campionamento
        &&$644$ $\mu$W (system)&&\\
        GC~\cite{GC5} & $>100$ kHz& $160\,\mu$W(trx), &$10$\,kb/s& %Epidural data tx
        Monitoring\\ 
        &&$365$ $\mu$W (tot)&&\\
        GC~\cite{GC7} & $10$ kHz & few $\mu$W (tx)&few kb/s&EMG data \\
        &&&&\\

        \hline
    \end{tabular}\label{TabTechsExp}
\end{table*}

\subsection{Comparison of intra-body testbeds for nervous system applications}

Table~\ref{TabTechsExp} compares the existing experimental solutions that exploit intra-body technologies for nervous system applications. While all the technologies are able to reach millimeter size, US already achieve sub-mm implementation. Coupling methods and US are the techniques that ensure lower power consumption (around $\mu$W), as shown by the parameters measuring the power, i.e., transmitted power, transceiver consumption, recording front-end consumption, and overall consumption.
These novel communication technologies exhibit great potential to be employed in ICT-based solutions to overcome nervous system pathologies, nevertheless, depending on the target application, one technology may be preferable to the other one. For example, the very small wavelength of light (hundreds of nanometers) allows optical-based methods to interact at the nanoscale with biological systems (e.g., proteins) to work at a single neuron level and even on genome, which is unfeasible with traditional electrical simulation employing electrodes. However, it is required highly complex and power consuming hardware for communication at THz frequencies and many new communication strategies need to be applied to optimize transmission and communication depending on the specific features of the employed hardware~\cite{Li2016}.
On the other hand, coupling methods working at low electromagnetic frequencies are able to cover longer distances inside the body since they exhibit lower signal attenuation compared with high frequency solutions. Therefore, they may be more suitable for data transfer from implants to a wearable data collector for further external remote processing and analysis through an IoMT infrastructure.

\vspace{-0.2cm}
\section{Future Directions}
\label{Directions}
\begin{figure}[!t]
\centering
	\includegraphics[width=0.45\textwidth]{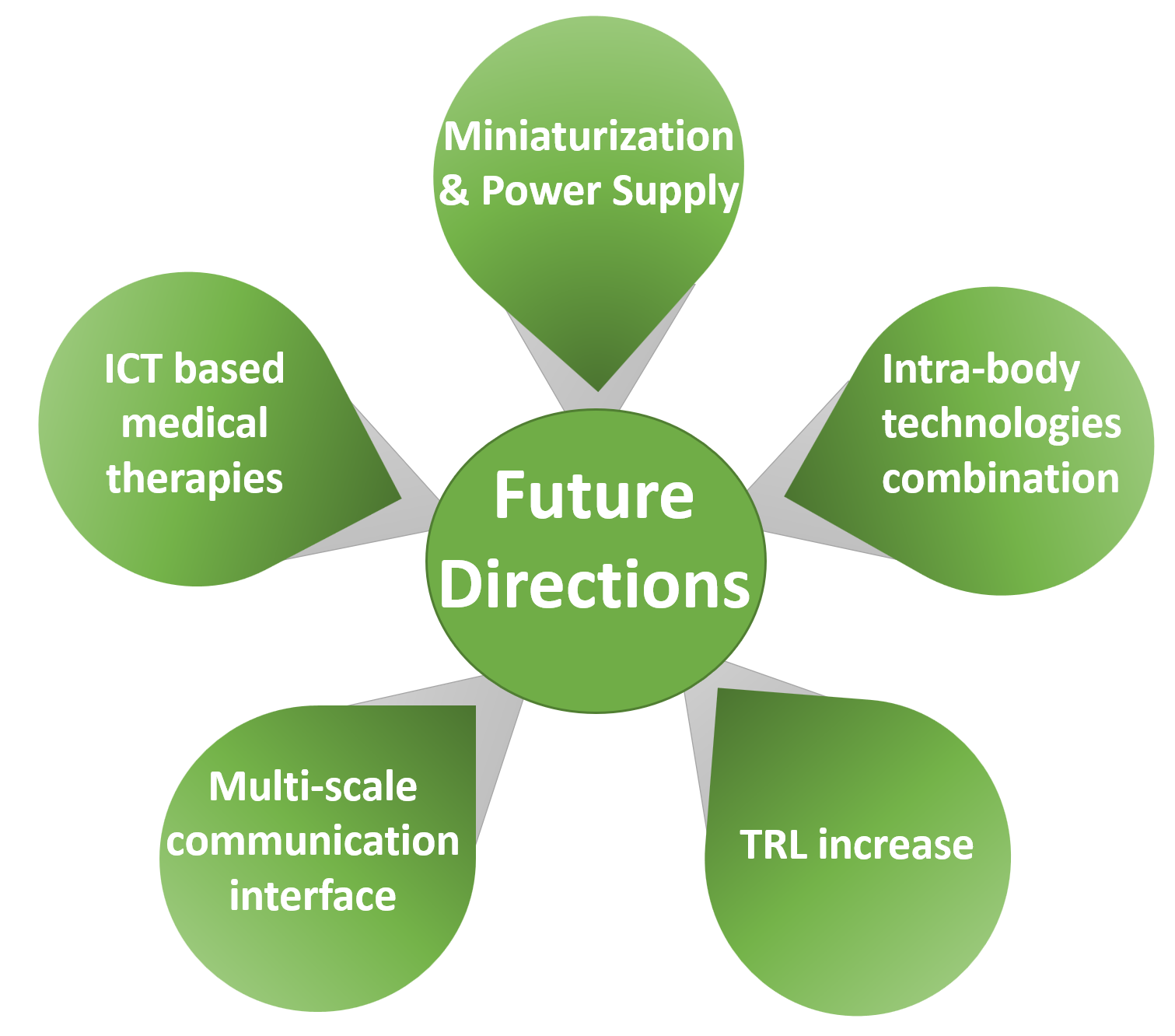}
	\caption{Summary of future directions in research and development.} \label{fig:FutureDirections}
\end{figure}
% \textcolor{red}{(check sopratutto challenges paper \cite{MC1}, \cite{OPTO}, \cite{Das2020} e poi vedi altre ref)}\\
\textcolor{black}{Although intra-body communication technologies have great potential for future 5G/B5G health applications, still several challenges need to be faced (see Fig.~\ref{fig:FutureDirections}).} In the following subsections we detail each of these challenging aspects.

\subsection{Miniaturization and Power Supply}%\textbf{Miniaturization and Power Supply:}
Current intra-body technologies are still on the mm scale and further miniaturization is required for long term implantation in human beings. In particular, research on opto miniaturization is just at the beginning~\cite{OPTO}, with developed devices in the mm size~\cite{Montgomery2015,Park2015}, while nano-devices are only at the design stage~\cite{OPTO3}. \textcolor{black}{However, ultrasonic solutions are already able to achieve sub-mm size ($0.8\,$mm$^{3}$), although some challenges still emerge. They are associated to circuit design complexity and limitations coming from the form factor associated to the use of ultrasonic frequencies.} Moreover, ultrasonic wireless systems are capable of moderate/high data rate (up to tens of Mb/s), while larger bandwidth would be required for advanced applications.  

%The power transfer to implants is another major task. The currently employed method is inductive coupling, suitable for under the skin devices with a cm size. However, for deep-implanted miniaturized devices with a size smaller than cm, alternatives should be explored such as ultrasounds harvesting 
Designing wireless sub-mm scale IMDs with cm-deep operation range presents power delivery and data transmission challenges. In terms of power transfer, IC~\cite{Ibrahim2018} is the conventional method adopted in biomedical implants in the size of cm and placed under the skin because of high power transmission efficiency and minimal dissipation of the magnetic field inside the tissue. However, miniaturization of implants to mm and even sub-mm dimensions, is difficult, hence alternatives should be explored. Among all power transfer systems, only US waves showed the capability to enable simultaneous power and data transfer in deep-implanted ($>2$ cm) mm-sized devices~\cite{Das2020}. 
Currently, ongoing projects~\cite{brain28nm} are exploring the idea of manufacturing novel systems using 28nm CMOS technology while including in a single integrated circuit  the analog front-end, the digital signal processing equipment and the energy harvesting subsystem, exploiting ultrasonic waves for both data communication and harvesting.

\subsection{Intra-body Technologies Combination} %\textbf{Intra-body Technologies Combination:} 
\label{CombTechs}
Since each of the aforementioned technologies exhibits specific features, an integration of them could be suitable to support complex medical applications. As an example, ultrasonic signals can be attenuated by the skull in case of brain applications or in general by bones, requiring an intermediate transceiver device, possibly exploiting coupling techniques operating at low EM frequencies under the skull to behave as gateway. 

A few solutions have been proposed in this direction. As an example~\cite{OPTO3}  designs the so called wi-opt neural dust device for deep brain stimulation, where ultrasonic vibrations are envisioned to harvest energy to power such nano-devices, and wireless optogenetics is used for the communication and stimulation on genetically engineered neurons using light, by exploiting the short wavelengths of optical frequencies able to target single neurons. 
The envisioned architecture consists of wi-opt neural dust devices that are embedded in various parts of the cortex and interface to neurons. Its main advantages are hence long-term deployment and the design of miniaturized devices that can self-generate power, besides the precise stimulation at single neuron level, fundamental for treating neurological diseases.

\begin{figure*}[!t]
    \centering
	\includegraphics[width=0.7\textwidth]{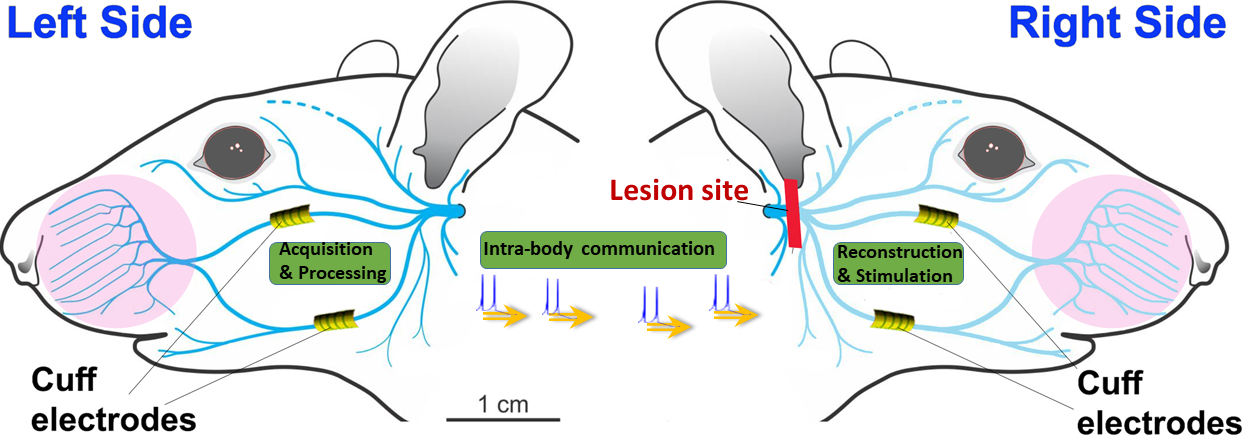}
	\caption{Future technologies combination: proposed setup for dual recording and stimulation in a rat with a nerve lesion to restore symmetrical facial movements via GC/US tissues transmission.} \label{fig:Rat}
\end{figure*}

Also, another combined solution has been recently proposed in our previous work~\cite{MagazineAll} to mix the sub-mm communication capability of US with the long distance one of GC to overcome a facial palsy, provided that both technologies exhibit high power efficiency and low invasivity.
An ideal implantable device should combine monitoring, recording, and stimulation mechanisms. Hence,
we design a dual neural signal acquisition and stimulation system, which opens the way to advanced and complex engineered solutions. Indeed, the  solution can be adapted to respond also to other peripheral nervous system diseases.
Specifically, we focus on the facial paralysis due to a nerve lesion, a common pathology that affects 1.81$\%$ of people, for which surgical corrections still represents the main therapeutic approach but adds 
nerve sacrifices and non negligible morbidity. Since most facial movements are symmetric, we devise an engineered implantable system to read the neural signal from a healthy facial nerve and transfer it to the contralateral, injured one using tissues as the propagation medium (Fig.~\ref{fig:Rat}) by means of GC and US. The stimulating signals are detected through cuff electrodes from the healthy nerve branches, processed and transmitted, so that they can be reproduced symmetrically at the level of the corresponding nerve branches on the pathological side, by the twin nerve stimulators. A preliminary version of the proposed architecture is presented in~\cite{MagazineAll} and we are currently testing the possibility of using either a direct link from one face side to the other one, or multi-hopping to split the path in multiple parts combining GC and US, to cope with the specific properties of the facial tissues that challenge the wireless intra-body transmission.

\textcolor{black}{The development of nerve interfaces has paved the way to the need to have more reliable means to communicate with the peripheral nervous system. This has led to a vast multi-disciplinary effort, to jointly solve surgical, medical, engineering, and software challenges while catalyzing rapidly maturing technologies~\cite{scholten2020interfacing}.}
Further studies need to be conducted, especially for complex medical applications, where the combination of intra-body technologies may play a crucial role to perform several tasks while guaranteeing multiple concurrent stringent requirements.

\subsection{TRL increase} %\textbf{TRL increase:} \textcolor{red}{(ANNA: check PRIN se più lungo)}
CC, GC, and US technologies have reached a degree of maturity that is on the verge to move from a proof of concept stage to a technological development. The development of these miniaturized transceivers is currently classified at TRL 4 being tested in the laboratory, and, for US, initial development of devices has already started. Further studies will boost the development of the manufacturing processes to build powerful miniaturized wireless communication devices. 
%Another important economic aspect, with a high impact on the public spending of a country, is represented by the saving from expensive medical cures. 

\subsection{Multiscale Communication Interface}

%\textcolor{red}{@TUTTI: secondo ME NON è CHIARO CHE BCI E BMI INTERAGISCONO DIRETTAMENTE CON LE CELLULE E NON RIGUARDANO ASPETTI DI COMUNICAIZONE. INOLTRE LA SEZIONE LA CHIAMEREI MULTISCALE COMMUNICATION INTERFACE COME NELLA FIGURA 6. HO GIA' CAMBIATO IL TUTOLO PERCHE' INTERFACING CON MOLECULAR COMMUNICATIONS NON C'ENTRA. FORSE PARLARE DI NEUROSCIENZE SEPPURE MOLTO INTERESSANTE IN QUESTA SEZIONE NON CI STA }
%\subsection{Multi-scale communication interface} %\textbf{Multi-scale communication interface:} 
More complex solutions could involve interfacing MC-based systems with other engineered solutions, which may imply the development of multi-scale communications depending on the considered intra-body technology. The paradigm of MC aims to develop artificial communication systems from biological components. 
%\textcolor{red}{da Sec. VII.B di \cite{IEEETMBMC2020-MC-THz} 09170555-TransMBMC-MCwithTHz.pdf:}
As an example, a biological transmitter may collect health
parameters and transmit them among the molecular nanonetwork, and a graphene based EM nano-device could be implanted into
the human body to deliver the information outside the
human body~\cite{IEEETMBMC2020-MC-THz}. This EM nano-device may be made up of a  chemical nanosensor, capable of detecting the concentration information coming from the MC nanonetworks and converting it to an electrical signal to be then transmitted trough the THz antenna embedded in the EM-nano device~\cite{IEEETMBMC2020-MC-THz}.

%\textcolor{red}{da Sec. 4.1 \cite{BioCyber2021} Review-Arxiv-Bio-CyberInterface-Sasi.pdf:}
Focusing on neural applications, brain computer interfaces (BCIs) and brain machine interfaces (BMIs) are interesting research areas to monitor and control the brain activity, conceived as a new treatment  or therapy for neural diseases, and cognitive problems. 
BCIs and BMIs refer to neurotechnologies capable to
observe the activity within the brain and decode or decipher
this to extract useful information~\cite{BMI1}. \textcolor{black}{They interface cells for signal acquisition using different methods, such as electrical or chemical based techniques, and future BMIs could be wirelessly interconnected with external devices exploiting intra-body communication methods. Some preliminary solutions have been developed in this direction~\cite{BioCyber2021}.}  
%Implanted devices that are able to translate the natural signals from brain into understandable machine commands, i.e., BMIs, have been developed in the past years~\cite{BioCyber2021}.

Recent studies in motor skills, such as handwriting, and speech from human cortical activity, along with the technology
advancements, enabled the observation of more channels of neural activity, which led to the new concepts for centralised/ distributed implant architectures. 
This calls for novel mechanisms for wireless powering, data transfer, new data processing, as well as more flexible substrates, miniaturized packaging, and surgical workflows. In particular, distributed BMIs have gained attention in the last four years. This brings new challenges to obtain an efficient power/data transfer and processing from multiple sites. As mentioned before, \textcolor{black}{US is a possible solution for joint powering and communication, indeed data communication can be achieved using the wireless power link to backscatter data},~\cite{BMI1},\textcolor{black}{~\cite{R2}},~\cite{Alesii}. Also, the latest advances in data processing, exploiting big data, machine and deep learning, are new opportunity for neural data compression~\cite{BMI1}.

In the future, these devices will be interconnected among them by means of intra-body communication technologies and even to the Internet, and may become bio-cyber interfaces to enable the exchange of any molecular signal among different intra-body networks, leading to the paradigm of the Internet of Bio-Nano Things (IoBNT). 
In IoBNT,  to interface the artificial systems based on MC to the Internet, other intra-body communication technologies, such as wireless optogenetics, can interact with both natural and artificial neurons based on MC, acting as a bio-cyber interface.
%and a unit based on one or more other intra-body communication technologies. 
As an example, a wireless optogenetic unit can act as a bio-cyber interface that enters information into the brain~\cite{OPTO}. Specifically, the bit transmission could be obtained through light stimulation of neuron that releases the vesicles to communicate to the post-synaptic neuron, whose mechanisms may be controlled through MC communication.

%Specific challenges related to MC applied to human health include incomplete biological knowledge, individuality,  variations, mutations and evolution.

%An exciting research field consists in the intersection of neurosciences and wireless communications, together with signal processing, computer science, and control theory~\cite{BMI2}. 
%On the one hand, neurosciences offer new applications to wireless networks, on the other hand wireless communication theory and next-generation wireless systems can provide new ways to study the brain.
%Studies in the first area lead to a new paradigm for a future wireless communications system: the brain-type communication, which differs from existing human- and machine-type communications, to develop new wireless services for networks with brain-in-the-loop~\cite{BMI2}.
%On the other side, the focus of the second area is related to how future wireless technologies can be beneficial for new research in neurosciences, which may include novel wireless-enabled BMIs and IoBNT, and information- and communication-theoretic methods of evaluating brain communications based on their chaotic nature~\cite{BMI2}.

\subsection{ICT-based medical therapies} %\textbf{ICT-based medical therapies:} 
Considering a long-term perspective, all these intra-body communication techniques could be extended to closed loop artificial neural systems that are completely implantable, made up of sensing nano-devices, transceivers and actuators capable to support motor recovery of paralized limbs, i.e. artificial prosthesis, or to bypass spinal injuries overcoming the current need for external processing. 

Also, blood flow monitoring is fundamental for treating some neurodegenerative disorders. Indeed, cerebral blood flow regulation is necessary for regular brain function, and a blood flow interruption may cause irreversible damage to neurons. In absence of a regular cerebral blood flow, defect in brain's functioning may occur, with consequent neurovascular dysfunction and conditions including Alzheimer’s disease~\cite{Nelson2020}. Nano-sensors integrated with intra-body communication technologies can be conceived to monitor high-risk people to prevent such neural damages.

%\textcolor{red}{ANNA: aggiungere BCI + intra-body comm da Sec. III.D \cite{IEEETMBMC2020}.}
Moreover, advanced BCI solutions could be envisioned with the support of intra-body technologies for the communication between the brain and the computer for several applications, from the usage of artificial intelligence to translate the brain signal to speech, to brain-computer
system based on augmented reality glasses where the system can read the word and phrase elaborated directly in the brain~\cite{IEEETMBMC2020}.

%\textcolor{red}{testo verde p. 7/14 di \cite{GC2}:}
Another exciting future application is a network of injectable, nano wireless neural implants. Their small size and wireless property allows
a complete freedom in choosing the locations of neural recording sites.
Since most neurological illnesses affect several brain regions, being able to monitor neural activity in multiple locations and observe intra-region communication is fundamental for a better understanding of dysfunctions. 
Also, as an example, placing multiple neural recording nano-implants in and around the focus of seizure activity could be advantageous for surgical planning or to monitor epileptic patients~\cite{GC2}.

The intra-body communication paradigm enables the communication of implants creating an intranet inside the body that may communicate outside with an IoT architecture for accomplishing more complex tasks, thus leading to emergence of novel ICT-based solutions.
% {\bf ABBIAMO DETTO TOGLIERE QUESTA ULTIMA FRASE}
% Finally, while this is not in the scope of this paper, we recognize the importance of a debate in view of an ethical regulation in this research area, as well as in the future resulting technologies.

\section{Conclusions}
\label{Conclusions}
In this paper we presented some promising intra-body communication technologies, application scenarios to the complex nervous system, and a number of testbeds underlying their applicability to some specific neural diseases. Furthermore, we discussed the main challenges that need to be faced for intra-body communication technologies, including multi-scale communication with combined technologies, miniaturization and power supply, and ICT-based solutions to treat neural diseases. We foresee that intra-body communication is a promising research area that may lead to the development of new multidisciplinary solutions to be applied in the neurological medical field. 

\section{Acknowledgements}
\begin{small}{
The work of LG was partially supported by the Ministero dell' Istruzione e della Ricerca, under the Project 4FRAILTY: ”Sensoristica intelligente, infrastrutture e modelli gestionali per la sicurezza di soggetti fragili” (ARS01$\_$00345).}
\end{small}
\appendix

\bibliographystyle{elsarticle-num} 
\bibliography{cas-refs}

\end{document}